\documentclass[prb,twocolumn,showpacs,amsmath,amssymb,superscriptaddress]{revtex4-2}

\usepackage{amsmath,amssymb,amsfonts,bm,color,graphicx,tabularx}

\usepackage[unicode=true,colorlinks=true]{hyperref}

\usepackage[etex=true,export]{adjustbox}
\usepackage{makecell}
\usepackage{multirow}
\usepackage{amsmath}
\usepackage{verbatim}

\hypersetup{linkcolor=blue, citecolor=blue,urlcolor=blue}
\usepackage{tabularx,multirow,array,diagbox}
\usepackage{adjustbox}

\newcommand{\beq}{\begin{equation}}
\newcommand{\eeq}{\end{equation}}
\newcommand{\beql}{\begin{equation*}}
\newcommand{\eeql}{\end{equation*}}
\newcommand{\beqn}{\begin{eqnarray}}
\newcommand{\eeqn}{\end{eqnarray}}

\begin{document}
\title{Surface-dependent Majorana vortex phases in topological crystalline insulators   }

\author{Xun-Jiang Luo}
\affiliation{School of Physics and Technology, Wuhan University, Wuhan 430072, China}
\affiliation{Department of Physics, Hong Kong University of Science and Technology, Clear Water Bay, 999077 Hong Kong, China}
\author{Xiao-Hong Pan}
\affiliation{Department of Physics, College of Physics, Optoelectronic Engineering, Jinan University, Guangzhou 510632, China}
\author{Yilin Shi}
\affiliation{School of Physics and Technology, Wuhan University, Wuhan 430072, China}
\author{Fengcheng Wu}
\email{wufcheng@whu.edu.cn}
\affiliation{School of Physics and Technology, Wuhan University, Wuhan 430072, China}
\affiliation{Wuhan Institute of Quantum Technology, Wuhan 430206, China}

\begin{abstract}
The topological crystalline insulator SnTe exhibits surface-dependent Dirac cones, which are located at non-time-reversal-invariant momenta on the (001) and (110) surfaces, but at time-reversal-invariant momenta on the (111) surface.
Motivated by the recent experimental evidence of Majorana vortex end modes (MVEMs) and their hybridization on the (001) surface [Nature $\bold{633}$, 71 (2024)], we present a comprehensive investigation of Majorana vortex phases in SnTe with proximity-induced superconductivity, including topological classification,  surface-state Hamiltonians analysis, and lattice model calculations. By utilizing rotational and magnetic mirror symmetries, we present two equivalent methods to reveal the topology of Majorana phases on different surfaces. We find that the MVEMs on the (001) and (110) surfaces are protected by both magnetic group and rotational symmetries. In contrast, the MVEMs on the (111) surface are protected by magnetic group or particle-hole symmetry. Due to the different properties of Dirac fermions in the $\bar{\Gamma}$ and $\bar{M}$ valleys on the (111) surfaces, including Fermi velocities and energy levels,  we find that abundant vortex phase transitions can occur for the [111]-direction vortex.  As the chemical potential shifts from the surface to bulk states, the number of robust MVEMs can change from $2\rightarrow 1\rightarrow 0$. These vortex transitions are characterized by both $Z$ winding number and $Z_2$ pfaffian topological invariants.

\end{abstract}
\maketitle

\section{Introduction}
Since the discovery of time-reversal invariant topological insulators (TIs) \cite{Kane2005,Kane2005a,Koenig2007}, topological phases of matter have emerged as one of the 
most exciting frontiers in condensed matter physics \cite{Hasan2010,Qi2011}.
In particular, topological superconductors (TSCs), which host Majorana zero modes (MZMs) satisfying non-Abelian statistics, have attracted extensive research interest due to their potential applications in topological quantum computation \cite{Nayak2008,Chiu2016,LuoMZM2024}. TSCs have been studied in various systems, including superconducting-proximitized topological insulators \cite{Fu2008,Fu2009,Pan2019,luo2018,Luo20241}, superconductor-semiconductor heterostructure \cite{Lutchyn2010,Oreg2010,Sau2010},  and superconducting vortex systems \cite{Fu2008,Hosur2011,Xu2016,Wang2018b,lichuang2022,2024arXiv241219096Y,zhang2025double}. In these systems, zero-dimensional unpaired MZMs are typically protected by particle-hole symmetry and possess a $Z_2$ topological classification \cite{Chiu2016}. On the other hand, crystalline symmetries can enrich the topological classification \cite{Fu2011c,Chiu2013,Liu2014} and lead to the emergence of topological crystalline insulators (TCIs) and superconductors \cite{Hsieh2012,Dziawa2012,Zhangfan2013,Teo2013a,Ueno2013,Hsieh2014,Shiozaki2014,Fang2014,xiongjun2014,Jadaun2013,Liu2015a,fangchen2015a,Kruthoff2017,Fang2019a,Cornfeld2019,YangHao2020,Song2020,PhysRevB.106.L121108,LiuTengteng2024,zhang2024fermi,Yamazaki2024,zhang2024topological}. Unlike conventional TIs, gapless boundary states in TCIs only appear at terminations that preserve specific crystal symmetries, which endows them with boundary-termination-dependent surface states. 

\begin{figure}[t]
\centering
\includegraphics[width=3.3in]{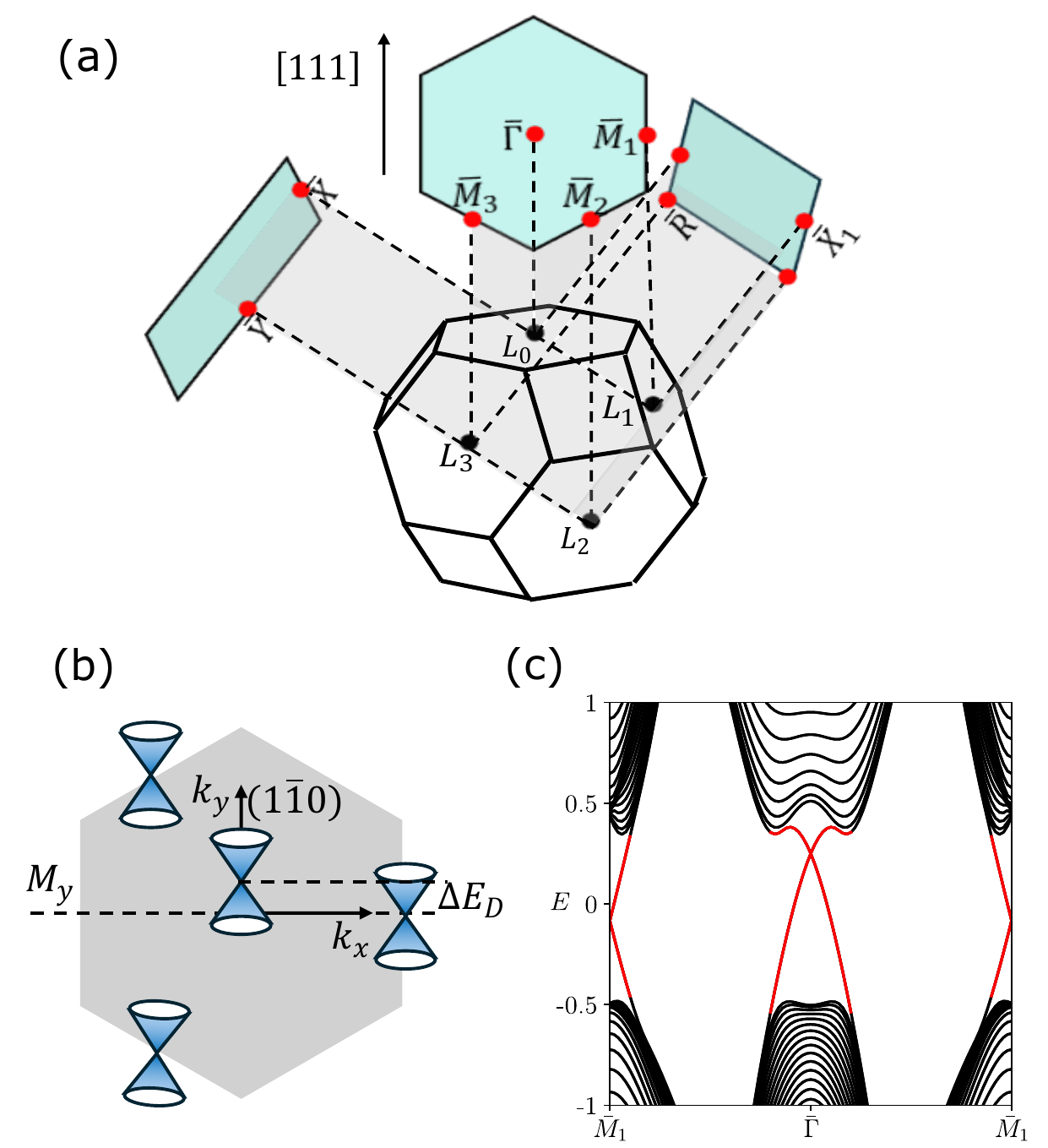}
\caption{(a) Bulk Brillouin zone of SnTe
and its projection onto the (001), (111), and (110) surfaces. The band inversion at $L$ points and their projection onto the surface Brillouin zone are illustrated. (b) The surface Dirac cones on the (111) surface. $\Delta E_{D}$ denotes the energy differences between the $\bar{\Gamma}$ and $\bar{M}$ valleys. (c) The energy band of the four-band model under the open boundary condition along the [111] direction. The red lines highlight the surface Dirac cones. }
\label{fig1}
\end{figure}

A notable example of this phenomenon is the mirror-protected TCI SnTe and related alloys
Pb$_x$Sn$_{1-x}$ (Te,Se), which have been extensively studied in both theory and experiment \cite{Tanaka2012,Xusuyang2012,Tanaka2013a,Yoshinori2013,fangchen2013,YanChenhui2014,Liujunwei2014,WangJianfeng2014,Zeljkovic2015,liujunwei2015b,KaiChang2016,Lau2019,YangHao2019,Liu2024,wangyuchun2024}. 
SnTe has band inversion at four $L$ points in the bulk [see Fig.~\ref{fig1}(a)], which gives rise to two types of surface Dirac cones. One is located at non-time-reversal-invariant momenta on the (001) and (110) surfaces and the other is located at time-reversal-invariant momenta on the (111) surface \cite{liujunwei2013}.  
It was theoretically predicted that when introducing a superconducting vortex on the (001) surface, there are two robust Majorana vortex end modes (MVEMs) which are protected by magnetic mirror symmetry \cite{Fang2014}. Another work \cite{xiongjun2014} demonstrated that these two MVEMs are protected by rotational symmetry. The two works \cite{Fang2014,xiongjun2014} confirm that the Majorana phase for the [001]-direction vortex belongs to a topological crystalline superconductor. Recently, significant progress has been made in this system, as experiments have reported the signatures of MVEMs in SnTe with proximity-induced superconductivity and moreover, their hybridization when the magnetic mirror symmetry is broken by in-plane magnetic fields \cite{YangHao2020,Liu2024}. This experiment not only provides strong evidence for the realization of topological crystalline superconductors but also introduces a field-tunable method to manipulate the fusion of a pair of MZMs. However, the exploration of Majorana phases for the [111]-direction vortex remains limited. It is noted that the (001) and (111) surfaces respect $C_{4v}$ and $C_{3v}$ point group symmetries, respectively. These point groups lead to distinct topological classifications of MVEMs \cite{Kobayashi2020}, and we summarize them in Table~\ref{tabcl}. 
On the (111) surface, there are four Dirac cones, with one at $\bar{\Gamma}$ valley and three at $\bar{M}$ valleys, where the latter three are related by a three-fold rotational symmetry but no symmetry connects the $\bar{\Gamma}$ and three $\bar{M}$ valleys. In a low-energy effective theory, the four Dirac cones can generate four MZMs in a vortex, but we show that only two of the four are symmetry protected, where one is contributed by the $\bar{\Gamma}$ valley and another is by the three $\bar{M}$ valleys. On the other hand,  the energy levels of the Dirac points located at $\bar{\Gamma}$ and three $\bar{M}$ points have an energy difference around $170$ meV informed by the angle-resolved photoemission spectroscopy results on the (111) surface \cite{Tanaka2013a}, as schematically
illustrated in Fig.~\ref{fig1}(b). This difference implies that 
there are abundant vortex phase transitions for the [111]-direction vortex by tuning chemical potential as no symmetry connects the $\bar{\Gamma}$ and $\bar{M}$ valleys.

In this work, we present a comprehensive investigation of the Majorana phases of different high-symmetry directional vortex in SnTe. Our study include topological classification, bulk and surface state Hamiltonians analysis, and lattice model calculations. By fully utilizing rotational and magnetic mirror symmetries, we provide two equivalent methods to reveal the topology of Majorana phases on different surfaces (the fourth and fifth rows in Table.~\ref{tabcl}).
We find that the two robust MVEMs on (001) and (110) surfaces are protected by both magnetic mirror and  rotational symmetries. 
Building on the surface Hamiltonian analysis, we show that two robust MVEMs can emerge on the (111) surface. One is contributed by the Dirac cone centered at the $\bar{\Gamma}$ point. 
Another is produced by a linear superposition of the three MZMs contributed by the surface Dirac cones located at the $\bar{M}_{1,2,3}$ points. Both of them have zero angular momentum and therefore, cannot be protected by the rotational symmetry,  but can be protected by the chiral symmetry generated by the magnetic mirror symmetry and particle-hole symmetry. 
Furthermore, we use both bulk low-energy effective Hamiltonian (Dirac model) at $L$ points and lattice models to investigate the vortex phase transitions for the [111]-direction vortex. We find that abundant vortex phase transitions can occur.  As the chemical potential shifts from the surface to bulk states, the number of robust MVEMs can change from $2\rightarrow 1\rightarrow 0$. These phase transitions can be characterized by both $Z$ winding number and $Z_2$ pfaffian topological invariants.

This paper is organized as follows. In Sec.~\ref{secII}, we present a comprehensive overview of the topological classification of MVEMs on the (001), (111), and (110) surfaces of SnTe and present two equivalent methods to fully reveal their topology. This classification is based on the effective Hamiltonian of surface Dirac cones. In Sec.~\ref{sec001}, we apply our theory to the Majorana phases on the (001)
and (110) surfaces. In Sec.~\ref{SecIII}, building on surface Hamiltonians analysis, we show that two robust MVEMs can emerge on the (111) surface and reveal their topology. In Sec.~\ref{SecV}, we use both bulk low-energy effective Hamiltonians and a four-band lattice model to study the vortex phase transitions of the [111]-direction vortex induced by the change of the chemical potential. In Sec.~\ref{SecV1}, we present a brief discussion and summary. Appendices \ref{windingnumberrelations}-\ref{Appendixc} complement the main text.

\begin{table}[b]
\centering
\setlength\tabcolsep{6pt}
\renewcommand{\multirowsetup}{\centering}
\renewcommand{\arraystretch}{1.2}
\caption{Classification of MVEMs on the (001), (111), and (110) surfaces of SnTe. The magnetic field is assumed to be perpendicular to the surface in each case. The BdG Hamiltonian $H_{\text{BdG}}$ respects the magnetic group symmetry $\mathcal{M_T}$, rotational symmetry $\mathcal{C}_n$, particle-hole symmetry $P$, and chiral symmetry $\mathcal{S}=\mathcal{M_T}P$. The winding number $W_{n}^{(p)}$ is calculated under the chiral symmetry $\mathcal{S}_n^{(p)}=\mathcal{C}_n^{p}\mathcal{S}$, with $0\leq p\leq n-1$ and $\mathcal{S}_n^{(0)}=\mathcal{S}$. The topological indices in the fourth and fifth rows provide two equivalent ways to characterize the topology of Majorana phases on different surfaces. The classification in this table is based on surface Dirac cones, assuming Fermi energy is within the bulk energy gap.}
\begin{tabular}{|c|c|c|c|}
\hline
surfaces&$(001)$&$(111)$ &$(110)$ \\
\hline
point group &$C_{4v}$&$C_{3v}$&$C_{2v}$\\
\hline
classification&$Z\times Z$&$Z$&$Z\times Z$\\
\hline
$w_{j'}$&
\makecell{$w_{0}=1$\\$w_{2}=1$}
& $w_{0}=2$
& \makecell{$w_{0}=1$\\$w_{1}=1$} \\ 
\hline
$W_{n}^{(p)}$ &\makecell{$W_{4}^{(0,2)}=2$\\ $W_{4}^{(1,3)}=0$} & $W_{3}^{(0,1,2)}=2$ & \makecell{$W_{2}^{(0)}=2$\\ $W_{2}^{(1)}=0$}\\
\hline
\end{tabular}
\label{tabcl}
\end{table}

\section{Vortex topology on different surfaces  }
\label{secII}
In a 3D Bogoliubov-de Gennes (BdG) system, a superconducting vortex along the $z$ direction breaks the in-plane translational symmetry. As a result, the entire system can be effectively treated as a 1D system. 
 In the Nambu basis $\Psi=\{c_{ k_{z}\alpha\uparrow},c_{ k_{z}\alpha\downarrow},c_{- k_{z}\alpha\uparrow}^{\dagger},c_{- k_{z}\alpha\downarrow}^{\dagger}\}^{T}$,
the BdG Hamiltonian for this effective 1D system can be expressed as,
 \beqn
H_{\mathrm{BdG}}(x,y,k_{z})=\left(\begin{array}{cc}
h\left(x,y,k_{z}\right)-\mu & \Delta(r, \theta) i s_y \\
-\Delta^*(r, \theta) i s_y & -h^T\left(x,y,-{k}_{z}\right)+\mu
\end{array}\right),\nonumber\\
\label{ha1}
\eeqn
where Pauli matrix $s_y$ acts on the spin space,  and the index $\alpha$ denotes the degrees of freedom associated with the orbitals and in-plane lattice sites. $h$ is the normal-state Hamiltonian and $\mu$ is the chemical potential. $\Delta(r,\theta)$ denotes the superconducting pairing potential and its specific form does not affect the topological classification. The particle-hole symmetry is given by $P=\tau_xK$, where $\tau_x$ denotes the Pauli matrix acting on particle-hole space and $K$ denotes complex conjugation.
$H_{\text{BdG}}$ belongs to the D symmetry class and has a $Z_2$ topological classification without considering crystalline symmetry \cite{Chiu2016}. 

\begin{figure}[t]
\centering
\includegraphics[width=3.3in]{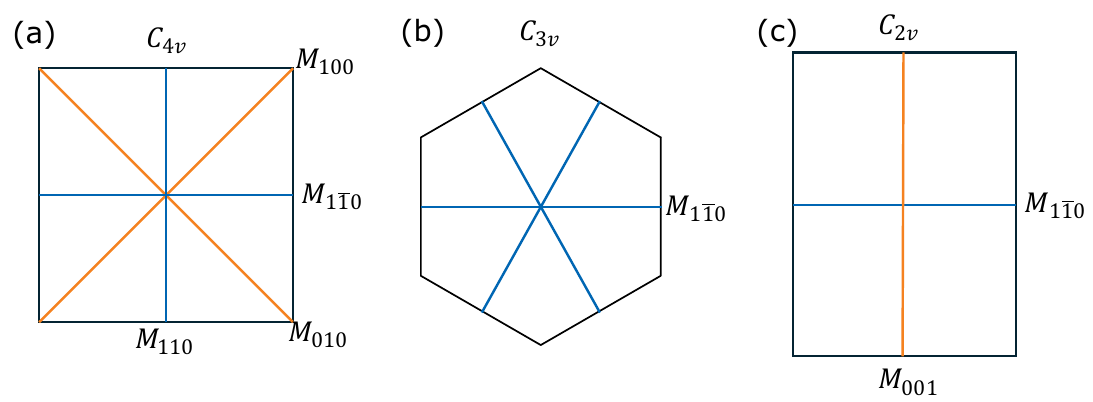}
\caption{Schematic illustration of mirror planes on (a) (001), (b) (111), and (c) (110) surfaces. In each panel, mirror planes labeled by the same (different) color generate identical (distinct) winding numbers, as presented in Eq.~\eqref{Wequi}.}
\label{figCn}
\end{figure}


 The crystal symmetry can enhance the topological classification of $H_{\text{BdG}}$, leading to the emergence of topological crystalline superconductors \cite{Kobayashi2020,Hu2023}. 
  In Eq.~\eqref{ha1},  when the $z$-axis is set to be the [001], [111], and [110] directions,  $h$ of describing SnTe respects the $C_{4v}$, $C_{3v}$, and $C_{2v}$ point group symmetry, respectively,  and time-reversal symmetry $T=is_yK$. The $C_{nv}$ ($n=2,3,4$) symmetries are generated by a rotation $C_n$ and vertical mirror operation $M_y$ that flips $y$ to $-y$. Throughout this paper, we choose $M_y$ as the mirror symmetry $M_{1\bar{1}0}$, which protects the topology of SnTe.

The superconducting vortex explicitly breaks both $T$ and $M_{y}$ symmetries, however, it preserves their combined operation, denoted by $\mathcal{M}_{\mathcal{T}}$,  with $(\mathcal{M}_{\mathcal{T}})^2={1}$. In this case, $H_{\text{BdG}}$ respects the chiral symmetry $\mathcal{S}=\mathcal{M}_{\mathcal{T}}P$, where $\{\mathcal{S},H_{\text{BdG}}\}=0$. Therefore, $H_{\text{BdG}}$ belongs to the BDI symmetry class and $\mathcal{M_T}$ symmetry enhances the topological classification of $H_{\text{BdG}}$ from $Z_2$ to $Z$.
By the chiral symmetry $\mathcal{S}$, we can define the winding number
\beqn
W=\frac{1}{4\pi i}\int_{0}^{2\pi}dk_z\text{Tr}[\mathcal{S}H_{\text{BdG}}^{-1}\partial_{k_z}H_{\text{BdG}}],
\label{wd}
\eeqn
which is formulated for Hamiltonian with periodic boundary condition along the $z$ direction.

In the presence of superconducting vortex, the correct form of rotational symmetry, denoted by $\mathcal{C}_n$,  is defined as \cite{xiongjun2014,Kobayashi2020}
 \beqn
 &\mathcal{C}_nc_{{k}_z\alpha s}^{\dagger}\mathcal{C}_n^{-1} = \sum_{\alpha^{\prime}s^{\prime} } c_{{k}_z\alpha^{\prime} s^{\prime}}^{\dagger}\left[C_n\right]_{\alpha^{\prime}s^{\prime}  ;\alpha s} e^{i \frac{\pi}{n}},\nonumber\\
&\mathcal{C}_nc_{{k}_z\alpha s}\mathcal{C}_n^{-1} = \sum_{\alpha^{\prime}s^{\prime} } \left[C_n\right]_{\alpha s;\alpha^{\prime}s^{\prime} }^{\dagger} e^{-i \frac{\pi}{n}}c_{{k}_z\alpha^{\prime} s^{\prime}}
 \eeqn
where $(\mathcal{C}_n)^{n}={1}$.
The rotational symmetry $\mathcal{C}_{n}$ enhances the topological classification of $H_{\text{BdG}}$ from $Z$ to $Z\times Z$ for both $n=4$ and $n=2$, but leaves the classification as $Z$ for $n=3$ \cite{Kobayashi2020}. Because $[\mathcal{C}_{n},H_{\mathrm{BdG}}]=0$, $H_{\mathrm{BdG}}$
 can be block-diagonalized into sectors spanned by eigenvectors of $\mathcal{C}_{n}$, namely,
 \beqn
 H_{\mathrm{BdG}}=H_{\mathrm{BdG}}^{(0)} \oplus H_{\mathrm{BdG}}^{(1)} \oplus \ldots \oplus H_{\mathrm{BdG}}^{(n-1)},
 \label{bh}
 \eeqn
where $ H_{\mathrm{BdG}}^{(j)}$ represents the Hamiltonian in the subsector with eigenvalue $e^{i 2 \pi j/ n}$ of $\mathcal{C}_n$, for $j=0,\cdots,n-1$. For the $j$th block associated with real $e^{i 2 \pi j/ n}$, $ H_{\mathrm{BdG}}^{(j)}$ respects the particle-hole symmetry $P$ and the chiral symmetry  $\mathcal{S}$ \cite{Kobayashi2020}, which belongs to the BDI symmetry class and has a $Z$ topological classification. For $n=4$, $3$, and $2$, respectively, the subsectors $j'=0,2$, $j'=0$, and $j'=0,1$ are associated with the real eigenvalue $e^{i 2 \pi j'/ n}$, which leads to the classification in the third row of Table.~\ref{tabcl}. For each block Hamiltonian $ H_{\mathrm{BdG}}^{(j')}$, we can define a winding number $w_{j'}$ by the chiral symmetry $\mathcal{S}$.
By bulk-boundary correspondence, $w_{j'}=n_{j',+}-n_{j',-}$, where  $n_{j',\pm}$ is the number MVEMs in open boundary condition with $\pm$ eigenvalue under the chiral symmetry  $\mathcal{S}$ in the $j'$th sector.
The number of robust MVEMs of the whole system is $\sum_{j'} |w_{j'}|$.

Because of  
$\{\mathcal{S},H_{\mathrm{BdG}}\}=0$ and $[\mathcal{C}_n,H_{\mathrm{BdG}}]=0$, the MVEMs can always be chosen as the eigenstates of $\mathcal{S}$ and $\mathcal{C}_n$. It is noted that $\mathcal{C}_n\mathcal{M_{T}}=\mathcal{M_{T}}\mathcal{C}_n^{-1}$ \cite{Kobayashi2020} and $[\mathcal{C}_n,P]=0$ \cite{fangcheng2017a}. Therefore, in the $j'$th block associated with real eigenvalue $e^{i 2 \pi j'/ n}$, we have $[\mathcal{C}_n,\mathcal{S}]=0$, which implies that the MVEMs contributed by these blocks can be chosen as the common eigenstates of $\mathcal{S}$ and $\mathcal{C}_n$. As $\{\mathcal{S},H_{\text{BdG}}\}=0$ and $[\mathcal{C}_n,H_{\text{BdG}}]=0$,  $\mathcal{S}$ and $\mathcal{C}_n$ symmetries protect the MVEMs that have the identical and different eigenvalues under $\mathcal{S}$ and $\mathcal{C}_n$, respectively. In Sec.~\ref{sec001}, we show that two  MVEMs on the (001) and (110) surface have the identical eigenvalue under $\mathcal{S}$ while have different eigenvalues under $\mathcal{C}_n$,  and therefore they are protected by both $\mathcal{S}$ and $\mathcal{C}_n$ symmetries. In Sec.~\ref{SecIII}, we show that two MVEMs on the (111) surface have the identical eigenvalue under both $\mathcal{S}$ and $\mathcal{C}_n$, and therefore they are protected by only $\mathcal{S}$ symmetry. We summarize our results for $w_{j'}$ in the fourth row of Table.~\ref{tabcl}.

By utilizing the rotational $\mathcal{C}_n$ and magnetic mirror $\mathcal{M_{T}}$ symmetries, the topology of $H_{\text{BdG}}$ can be alternatively revealed by a seris of winding numbers.  Because $[H_{\mathrm{BdG}},\mathcal{C}_n]=0$, 
we have 
\beqn
\{\mathcal{C}_n^{p}\mathcal{S},H_{\mathrm{BdG}}\}=\{\mathcal{C}_n^{p}\mathcal{M_{T}}P,H_{\mathrm{BdG}}\}=0,
\eeqn
where $0\leq p\leq n-1$. This implies that we can define four, three, and two chiral symmetries $\mathcal{S}_n^{(p)}=\mathcal{C}_n^{p}\mathcal{S}$ for $n=4$, $n=3$, and $n=2$, respectively, with $\mathcal{S}_n^{(0)}=\mathcal{S}$. By replacing $\mathcal{S}$ with $\mathcal{S}_n^{(p)}$ in Eq.~\eqref{wd}, we obtain $n$ winding numbers $W_{n}^{(0,\cdots,n-1)}$ which act as another way to fully characterize the topology of $H_{\text{BdG}}$.

However, not all the winding numbers $W_{n}^{(p)}$ are independent since some of the chiral symmetries can be related by unitary transformation. We can demonstrate that 
\begin{equation}
\begin{aligned}
&W_{4}^{(0)}=W_{4}^{(2)}, W_{4}^{(1)}=W_{4}^{(3)},\\
&W_{3}^{(0)}=W_{3}^{(1)}=W_{3}^{(2)}.
\end{aligned}
\label{Wequi}
\end{equation}
Because of $\mathcal{C}_n\mathcal{M_{T}}=\mathcal{M_{T}}\mathcal{C}_n^{-1}$ and $[\mathcal{C}_n,P]=0$, we have
\beqn
\mathcal{S}_{4}^{(i+2)}=\mathcal{C}_{4}^2\mathcal{S}_{4}^{(i)}=\mathcal{C}_{4}^{i+2}\mathcal{M_{T}}P=\mathcal{C}_{4}\mathcal{S}_{4}^{(i)}\mathcal{C}_{4}^{-1},
\eeqn
with $i=0,1$. Thus, the chiral symmetry $S_{4}^{(i+2)}$ and $S_{4}^{(i)}$ are related by a unitary transformation $\mathcal{C}_{4}$. For $n=4$, since $[\mathcal{C}_4,H_{\text{BdG}}(k_z)]=0$ at arbitrary $k_z$, we have $W_{4}^{(i)}=W_{4}^{(i+2)}$ according to Eq.~\eqref{wd}.
For $n=3$, because $\mathcal{C}_{3}^{3}=1$, we have $\mathcal{C}_3=(\mathcal{C}_3^{-1})^2$, which leads to
\beqn
\mathcal{S}_{3}^{(i+1)}=\mathcal{C}_{3}\mathcal{S}_{3}^{(i)}=\mathcal{C}_3^{-1}\mathcal{S}_{3}^{(i)}\mathcal{C}_{3},
\eeqn
with $i=0,1$. Therefore, we have $W_{3}^{(0)}=W_{3}^{(1)}=W_{3}^{(2)}$. 
For $n=2$, $W_{2}^{(0)}$ and $W_{2}^{(1)}$ are independent as no symmetry operator in the $C_{2v}$ point group relates the two.  We provide a schematic illustration of Eq.~\eqref{Wequi} in Fig.~\ref{figCn}.

The two sets of winding numbers given respectively by $w_{j^{\prime}}$ and $W_n^{(p)}$ are related to each other.
By using the definition of the winding number in Eq.~\eqref{wd} and the block-diagonalized Hamiltonian in Eq.~\eqref{bh}, it can be proved that (see Appendix~\ref{windingnumberrelations} for the proof),
\begin{equation}
\begin{aligned}
&W_n^{(0)}=\sum_{j^{\prime}}w_{j^{\prime}},\\
&W_{4}^{(1)}=w_0-w_{2},\\
&W_{2}^{(1)}=w_0-w_{1}.
\end{aligned}
\end{equation}
Thus,  the two sets of winding numbers provide an equivalent characterization. We emphasize that the numerical calculation of $W_{n}^{(p)}$ , without the need to extract the subsector Hamiltonian $H_{\text{BdG}}^{(j')}$, is more convenient than that of $w_{j^{\prime}}$. In the fifth row of Table.~\ref{tabcl}, we list the results on $W_{n}^{(p)}$
associated with different surfaces, which can be derived from both surface-state Hamiltonian (Sec.~\ref{SecIII})  and numerical calculations in the lattice model (Sec.~\ref{SecV}). The fourth and fifth rows in Table.~\ref{tabcl} provide two equivalent ways to characterize the topology of different directional vortex by using both rotational $\mathcal{C}_n$ and magnetic group $\mathcal{M_T}$ symmetries.

\section{Majorana phases on the (001) and (110) surfaces}
\label{sec001}
We now apply our theory to MVEMs on the (001) and (110) surfaces of SnTe with proximity-induced superconductivity. The Majorana vortex phase of SnTe on the (001) surface, corresponding to $n=4$, was studied in the works \cite{xiongjun2014,Fang2014,wangyuchun2024}. In Ref.~\cite{Fang2014}, the authors demonstrated that there are  two robust MZMs localized at a vortex core. These two MZMs are protected by the $\mathcal{M_{T}}$ symmetry, or the chiral symmetry $\mathcal{S}$.  In Ref.~\cite{wangyuchun2024}, the winding number of $H_{\text{BdG}}$ was calculated using a lattice model, revealing a value of 2, which further confirms the existence of two robust MZMs at a vortex.   In Ref.~\cite{xiongjun2014}, the authors presented a different classification theory from that of Ref.~\cite{Fang2014} and  showed that the two robust MZMs on the (001) surface are protected by the rotational symmetry $\mathcal{C}_4$. Moreover, although the $\mathcal{C}_4$ symmetry is broken by  an in-plane  field, the two MVEMs are restored exactly one time whenever the in-plane field varies $\pi/2$. The different classification theory presented in Ref.~\cite{Fang2014} and Ref.~\cite{xiongjun2014} can be understood from our unified perspective. As explained in Appendix \ref{appendixa}, from the linear superposition of the four MVEMs contributed by the four surface Dirac cones on the (001) surface, we can obtain two MVEMs ${\gamma}_{1}$ and ${\gamma}_{2}$, which are common eigenstates of $\mathcal{S}$ and $\mathcal{C}_4$ symmetries. Both ${\gamma}_{1}$ and ${\gamma}_{2}$ have the eigenvalue $1$ under $\mathcal{S}$, and they have the eigenvalue $1$ and $-1$, respectively, under $\mathcal{C}_4$. From these eigenvalue configurations, we can derive that $W_{4}^{(0)}=W_{4}^{(2)}=2$ and $W_{4}^{(1)}=W_{4}^{(3)}=0$, and therefore the results for different characterization methods are related. The winding number results can be further numerically verified by using a four-band lattice model (see Appendix~\ref{Appendixd}). Consequently,  the two MVEMs  ${\gamma}_{1}$ and ${\gamma}_{2}$ are protected by both the chiral symmetry $\mathcal{S}$ and rotational symmetry $\mathcal{C}_4$.  In other words, the two robust MVEMs are protected by the magnetic mirror symmetry with the mirror being $M_{110}$ or $M_{1\bar{1}0}$, rather than $M_{100}$ and $M_{010}$.
Although the $\mathcal{C}_4$ symmetry is broken by an in-plane field, the chiral symmetry associated with the mirror symmetry is restored if the field is applied along the [110] or [$1\bar{1}0$] direction. In this scenario, the two MZMs survive, which is consistent with the analysis in  Ref.~\cite{xiongjun2014}.

SnTe hosts two surface Dirac cones on the (110) surfaces \cite{liujunwei2013}. These two Dirac cones are located at the $M_{1\bar{1}0}$ invariant line in the surface Brillouin zone and are related by the $T$ symmetry \cite{liujunwei2013}. Through a similar surface Hamiltonian analysis as that for the (001) surface (see Appendix~\ref{Appendixb}), it can be shown that two MVEMs, denoted by $\gamma_3$ and $\gamma_4$, can be obtained and $\gamma_{3,4}$ can be chosen as the common eigenstates of $\mathcal{S}$ and $\mathcal{C}_2$ symmetries. $\gamma_3$ and $\gamma_4$ have the eigenvalue $1$ under $\mathcal{S}$, and they have the eigenvalue $1$ and $-1$, respectively, under $\mathcal{C}_2$, which leads to $W_2^{(0)}=2$ and $W_2^{(1)}=0$, and the results for different characterization methods are again related. $W_2^{(0,1)}$ can also be directly calculated by using a four-band lattice model (see Appendix~\ref{Appendixd}). 
The above discussions assume chemical potential $\mu$ at the surface Dirac point (i.e., $\mu=0$).
Numerically, we find that as $|\mu|$ increases, $W_{4}^{(0)}$ and  $W_{2}^{(0)}$ change from $2$ to $0$ (see Appendix.~\ref{Appendixd}) and no unpaired MVEMs appear on the (001) and (110) surfaces. This is consistent with the fact that all the Dirac cones on the (001) or (110) surfaces are related by crystalline symmetry or $T$ symmetry.

\section{Majorana phases on the (111) surface}
\label{SecIII}

SnTe hosts four Dirac cones on the (111) surface, which are located at the time-reversal-invariant $\bar{\Gamma}$ and  $\bar{M}_{1,2,3}$ points, respectively, as schematically illustrated in Fig.~\ref{fig1}(b). 
The surface Hamiltonian of describing the Dirac cone centered at $\bar{\Gamma}$ and $\bar{M}_{1,2,3}$ can be written as
\beqn
\hat{h}_{i=0,1,2,3}=\sum_{|\bm{q}|<\Lambda, s, s^{\prime}=\uparrow, \downarrow} h_{i}^{s s^{\prime}}(\bm{q}) f_{i s}^{\dagger}(\bm{q}) f_{i s^{\prime}}(\bm{q}),
\eeqn
where $f_{0 s}(\bm{q})$ and $f_{j s}(\bm{q})$ are the annihilation operator at $\bm k=\bar{\Gamma}+\bm{q}$ and $\bm k=\bar{M}_{j}+\bm{q}$, respectively, for $j=1,2,3$.   $h_0$ is fixed by choosing the representation of the little group at $\bar{\Gamma}$ to be $T=is_yK$,  $M_y=is_y$, and  $C_3=e^{i\pi s_z/3}$, which leads to $h_0(\bm q)=v_1(q_xs_y-q_ys_x)$. The little group at $\bar{M}_1$ is generated by the 
$T$ and $M_y$ symmetries. From the chosen representation, we can derive $h_1(\bm q)=v_2(q_xs_y-q_ys_x)$. In $h_{0,1}$, we assume isotropic Fermi velocity for analytical convenience. $v_1$ and $v_2$ share the same sign, as required by the mirror Chern number $C_{M}=2$ \cite{Hsieh2012}.
We note that the Dirac point energies at $\bar{\Gamma}$ and  $\bar{M}_{1,2,3}$ valleys are generally different, which is neglected in this section but studied in Sec.\ref{SecV}. 
Using the $C_3$ symmetry, we can fix the gauges for Dirac cones centered at $\bar{M}_{2,3}$ as $f_{2,3}\left(C_3 \bm{q}\right) \equiv$ $C_3 f_{1,2}(\bm{q}) C_3^{-1}$, which lead to 
\beqn
h_2(\bm q)=h_1(C_3^{-1}\bm q), \quad h_3(\bm q)=h_1(C_3\bm q),
\eeqn
where $C_3(q_x,q_y)\rightarrow (-q_x/2-\sqrt{3}q_y/2, \sqrt{3}q_x/2-q_y/2)$. In Table.~\ref{tf}, we  summarize the transformations of the annihilation operators $f_{is}$ ($i=0,1,2,3$) under the $C_{3v}$ and $T$ symmetries.

\begin{table}[t]
\centering
\setlength\tabcolsep{1.2pt}
\renewcommand{\multirowsetup}{\centering}
\renewcommand{\arraystretch}{1.8}
\caption{The transformations of the annihilation operators $f_{is}$ ($i=0,1,2,3$) under the $C_{3v}$ and $T$ symmetries. }
\begin{tabular}{|c|c|c|c|c|}
\hline
\diagbox[trim=l,height=3\line]{symmetries}{operators} &$f_0$& $f_1$ & $f_2$ & $f_3$ \\
\hline
$M_y$ & $\left(i s_y\right) f_0$ &$\left(i s_y\right) f_1$ & $\left(-i s_y\right) f_3$ & $\left(-i s_y\right) f_2$ \\
\hline
$C_3$ &$(e^{i\pi s_z/3}) f_0$ & $f_2$ & $f_3$ & $-f_1$ \\
\hline
$T$ &$(i s_y) f_0$ & $(i s_y) f_1$ & $(i s_y f_2$ & $(i s_y) f_3$ \\
\hline
\end{tabular}
\label{tf}
\end{table}

When considering proximity-induced $s$-wave pairing, the four surface Dirac cones favor intravalley superconducting pairing since they are located at time-reversal-invariant momenta.  The surface state Hamiltonian, incorporating the superconducting vortex, for each Dirac cone, can be expressed as,
 \beqn
\mathcal{H}_{i}(\bm q)=\left(\begin{array}{cc}
h_{i}(\bm q) & \Delta(r, \theta) i s_y \\
-\Delta^*(r, \theta) i s_y & -h_{i}^T\left(-\bm q\right)
\end{array}\right),
\label{ha}
\eeqn
where $i=0,1,2,3$, $\Delta (r, \theta) = \Delta(r) e^{-i \theta}$ with $\theta$ being the polar angle, and $\mu=0$ (i.e., the chemical potential is at the Dirac point in each valley). From each of these Hamiltonians,  a zero-energy Majorana vortex mode can be derived (see Appendix.~\ref{Appendixc}). In total, we obtain four MZMs, expressed as follows:
\begin{equation}
\begin{aligned}
&\gamma_0=(f_{0\downarrow}+f_{0\downarrow}^{\dagger})e^{-\int_{0}^{r}\Delta(r^{\prime})/v_1dr^{\prime}},\\
&\gamma_1=(f_{1\downarrow}+f_{1\downarrow}^{\dagger})e^{-\int_{0}^{r}\Delta(r^{\prime})/v_2dr^{\prime}},\\
&\gamma_2=(e^{-i\pi/3}f_{2\downarrow}+e^{i\pi/3}f_{2\downarrow}^{\dagger})e^{-\int_{0}^{r}\Delta(r^{\prime})/v_2dr^{\prime}},\\
&\gamma_3=(e^{-i2\pi/3}f_{3\downarrow}+e^{i2\pi/3}f_{3\downarrow}^{\dagger})e^{-\int_{0}^{r}\Delta(r^{\prime})/v_2dr^{\prime}},
\end{aligned}
\end{equation}
where we assume that $ \Delta(r) $ and $v_1$ share the identical sign. Although we obtain four MZMs,  a further symmetry analysis is  needed to assess the robustness of $\gamma_{0,1,2,3}$.

\begin{figure*}
\centering
\includegraphics[width=6.5in]{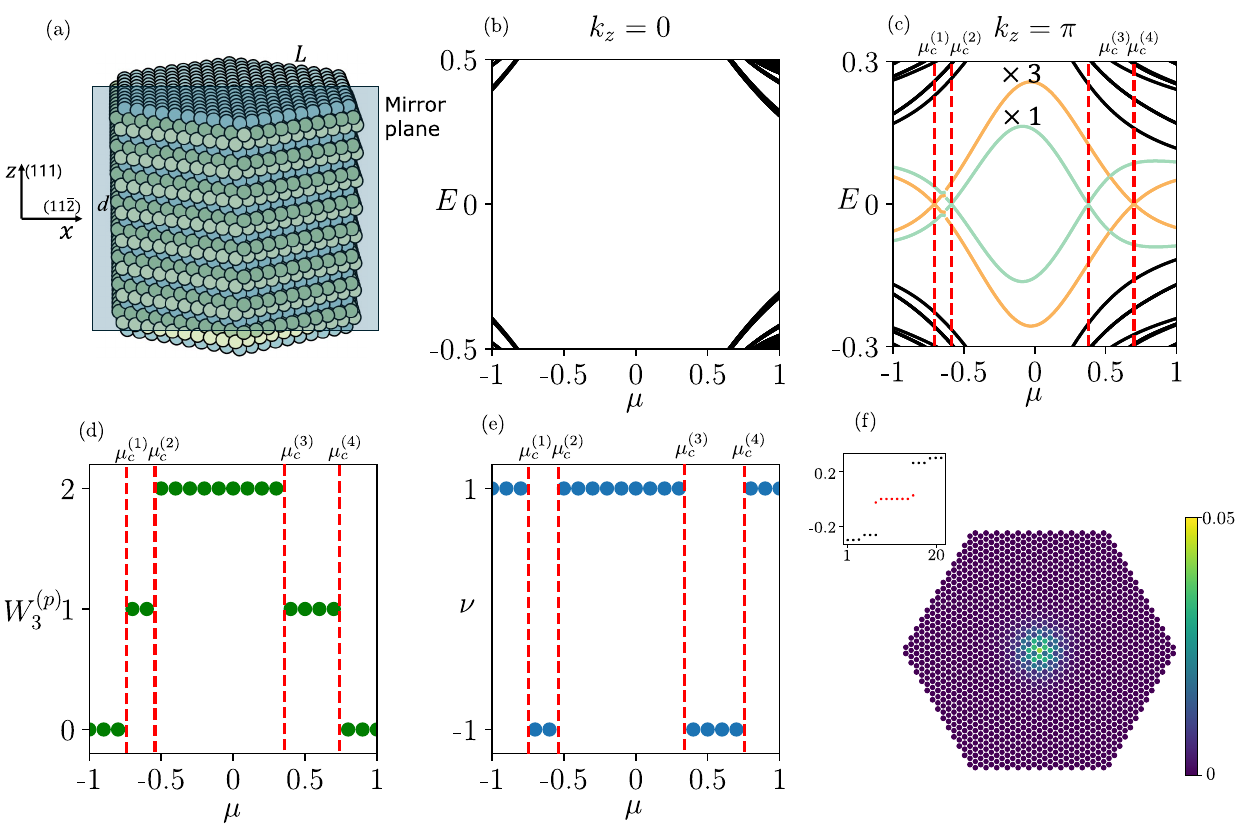}
\caption{(a) Schematic illustration of the ABC stacking system along the (111) direction. $d$ and $L$ denote the length of side of the (111) surface and height of the system, respectively. (b) and (c) The evolution of the energies with $\mu$ for a wire system with  $k_{z}=0$ and $k_{z}=\pi$, respectively. $k_z$ is in unit of $1/(2\sqrt{3}a)$. The orange bands have threefold degeneracies and the green bands have no extra degeneracy.  (d) and (e) The winding number $W_{3}^{(p)}$ and pfaffian topological invariant $\nu$ as functions of $\mu$, respectively.  (f) The real-space distribution of the eight states near the zero energy (see the inset) on the surfaces. In (b)-(f), $L=13\sqrt{2}a$. In (d)-(f), $d=12\sqrt{3}a$. We take $\mu=0$ in (f). The common parameters are taken as $\Delta_0=0.5$ and $\xi=2a$.
 } 
\label{fig2}
\end{figure*}

As elucidated in Sec.~\ref{secII}, the entire system hosts the chiral symmetry $\mathcal{S}$ and three-fold rotational symmetry $\mathcal{C}_3$, with their specific forms detailed in Appendix~\ref{Appendixc}. The $\mathcal{S}$ and  $\mathcal{C}_3$ symmetries act on $\gamma_{0,1,2,3}$ as
\begin{equation}
\begin{aligned}
&\mathcal{S}\gamma_{0,1}\mathcal{S}^{-1}=\gamma_{0,1}, \quad\mathcal{S}\gamma_{2,3}\mathcal{S}^{-1}=\gamma_{3,2},\\
&\mathcal{C}_3\gamma_{0}\mathcal{C}_3^{-1}=\gamma_{0}, \quad
\mathcal{C}_3\gamma_{1,2,3}\mathcal{C}_3^{-1}=\gamma_{2,3,1}.
\label{sc3}
\end{aligned}
\end{equation}
Consequently, in the zero-energy subspace expanded by $\gamma_{0,1,2,3}$, the representation of $\mathcal{S}$ and $\mathcal{C}_3$ are, respectively, given by,
\beqn
\mathcal{S}=\left(\begin{matrix}
1 & 0 & 0 & 0\\
0 & 1 & 0 & 0\\
0 & 0 & 0 & 1\\
0& 0 & 1 & 0
\end{matrix}\right),\quad
\mathcal{C}_3=\left(\begin{matrix}
1 & 0 & 0 & 0\\
0 & 0 & 1 & 0\\
0 & 0 & 0 & 1\\
0& 1 & 0 & 0
\end{matrix}\right).
\eeqn
Since the $\mathcal{S}$ symmetry only protects the MZMs that have the identical eigenvalues, the number of robust MZMs is
\beqn
W=\text{Tr}(\mathcal{S})=2,
\eeqn
 which implies that two MZMs out of four are protected by the $\mathcal{S}$ symmetry.
It is noted that both $\mathcal{C}_3\mathcal{S}$ and $\mathcal{C}_3^{2}\mathcal{S}$ also serve as the chiral symmetries of the system and we have 
\beqn
\text{Tr}(\mathcal{S})=\text{Tr}(\mathcal{C}_3\mathcal{S})=\text{Tr}(\mathcal{C}_3^2\mathcal{S})=2.
\eeqn
This indicates that $W_{3}^{(1)}=W_{3}^{(2)}=W_{3}^{(3)}=2$, which is consistent with the bulk calculation presented in Sec.~\ref{SecV}. The four zero-energy states can also be chosen as the eigenstates of $\mathcal{C}_3$ symmetry.
 By diagonalizing $\mathcal{C}_3$, we derive the MZMs with fixed angular momentum, yielding,
\begin{equation}
\begin{aligned}
&\tilde{\gamma}_{0}=\gamma_0, \tilde{\gamma}_1=(\gamma_1+\gamma_2+\gamma_3)/\sqrt{3},\\
&\tilde{\gamma}_2=(e^{i2\pi/3}\gamma_1+e^{-i2\pi/3}\gamma_2+\gamma_3)/\sqrt{3},\\
&\tilde{\gamma}_3=(e^{-i2\pi/3}\gamma_1+e^{i2\pi/3}\gamma_2+\gamma_3)/\sqrt{3},
\end{aligned}
\label{gamma0123}
\end{equation}
where $\tilde{\gamma}_{0,1}$, $\tilde{\gamma}_{2}$, and $\tilde{\gamma}_{3}$ have the angular momentum $J_z=0$, $J_z=-1$, and $J_z=1$, respectively. Furthermore, we have $\mathcal{S}\tilde{\gamma}_{0,1}\mathcal{S}^{-1}=\tilde{\gamma}_{0,1}$ and $\mathcal{S}\tilde{\gamma}_{2,3}\mathcal{S}^{-1}=\tilde{\gamma}_{3,2}$. Thus, the MZMs $\tilde{\gamma}_{0,1}$ are the common eigenstates of $\mathcal{S}$ and $\mathcal{C}_3$ and are protected by the chiral symmetry $\mathcal{S}$ while the $\mathcal{C}_3$ symmetry does not provide a protection. This is distinguished from the vortex physics on the (001) and (110) surfaces, where two robust MVEMs are protected by both the chiral and rotational symmetries.

\section{Vortex phase transitions}
\label{SecV}
On the (111) surface, the Dirac cones at $\bar{\Gamma}$ and  $\bar{M}_{1,2,3}$ valleys have different properties, for example, the energy levels and the Fermi velocities. As  no symmetry relates the $\bar{\Gamma}$ and  $\bar{M}_{1,2,3}$ valleys, MVEMs derived from these two different types of valleys are subjected to different critical chemical potentials. This suggests a rich pattern of vortex phase transitions tuned by the chemical potential for vortex along [111] direction.

The surface-state Hamiltonians analysis presented in Sec.~\ref{SecIII} captures the low-energy physics when the chemical potential $\mu$ is near the Dirac points, but it can not fully describe the vortex phase transition as $|\mu|$ increases.
This transition is essentially determined by the bulk states \cite{Hosur2011}. 
For the bulk low-energy states in SnTe, there are four time-reversal invariant $L$ points, including $L_{0,1,2,3}$ as illustrated in Fig.~\ref{fig1}(a). $L_{0}$ ($L_{1,2,3}$) points are projected onto the $\bar{\Gamma}$ ($\bar{M}_{1,2,3}$) points on the (111) surface Brillouin zone.
The bulk $k.p$ Hamiltonian at $L_{i}$ is described by an effective Dirac model
\cite{Mitchell1966,Adler1973,Hsieh2012}
\beqn
H(\mathbf{k})&=&(c_{||} (k_1^2+k_2^2)+c_3 k_3^2)\sigma_0s_0 \nonumber\\
&+&(m_0-\epsilon_{||} (k_1^2+k_2^2)-\epsilon_3 k_3^2)\sigma_zs_0 \nonumber\\
&+&v_3 k_3\sigma_y+v_{||}(k_1 s_y-k_2 s_x)\sigma_x,
\label{HkDirac}
\eeqn
where Pauli matrices $\sigma$ act on the orbital space, $k_3$ is along the $\Gamma L_i$ direction, $\{m_0, c_{||,3},\epsilon_{||,3}, v_{||,3}\}$ are model parameters. We note that $H(\mathbf{k})$ has a samilar form as the low-energy Hamiltonian of a 3D topological insulator\cite{Zhang2009}. 
We consider a magnetic field $\bm B$ along the  $\Gamma L_0$ direction, which is also the [111] direction. Then $\bm B$ is parallel to $k_3$ for the effective Hamiltonian in Eq.~\eqref{HkDirac} at $L_0$. We label the critical chemical potential of the vortex phase transition for the $L_0$ valley as $\mu_c^{L_0}$.
For the other three valleys of $L_{1,2,3}$, the $\bm B$ field is not parallel to the corresponding $k_3$ direction. Because of the anisotropy in the Dirac Hamiltonian in Eq.~\eqref{HkDirac}, the critical chemical potentials $\mu_c^{L_{1,2,3}}$ of the vortex phase transition for the $L_{1,2,3}$ valleys are generally different from $\mu_c^{L_0}$ for $\bm B$ along $\Gamma L_0$ direction. On the other hand, the threefold rotation around the [111] axis relates the $L_{1,2,3}$ valleys, and therefore, $\mu_c^{L_{1}}=\mu_c^{L_{2}}=\mu_c^{L_{3}}$.
From this analysis based on bulk states, we also expect two critical chemical potentials as $\mu$ increases (decreases) to enter the bulk conduction (valence) bands.

To investigate the $\mu$-tuned vortex phase transitions at a more quantitative level, a lattice model is needed. The band structure of SnTe can be described by a twelve-band tight-binding model that incorporates $p_{x,y,z}$ orbitals\cite{Hsieh2012,wangyuchun2024}. To mitigate the complexity of calculations while still capturing the essential physics, we use a simplified four-band model developed in Ref.~\cite{Fang2014} in our calculations. 
The Bloch Hamiltonian of the four-band model  is \cite{Fang2014}
\begin{equation}
\begin{aligned}
h(\mathbf{k}) & =[m-t_1(\cos 2 k_1a+\cos  2k_2a+\cos 2k_3a)] \sigma_{z}s_0 \\
&+t_2[\sin k_1a\left(\cos k_2a+\cos k_3a\right) \sigma_{x }s_x \\
&+\sin k_2a(\cos k_1a+\cos k_3a) \sigma_{x}s_y\\
&+\sin k_3a(\cos k_1a+\cos k_2a) \sigma_{x}s_z]\\
&+t_3[\cos k_1a(\cos k_2a+\cos k_3a)+\cos k_2a\cos k_3a]\sigma_0s_0,
\label{fourband}
\end{aligned}
\end{equation}
where the Pauli matrices $\sigma$ and $s$ act on the orbital and spin space, respectively.
We choose the model parameters as $(m, t_1, t_2,t_3)=(2.5,-1,1,1)$.  Here the $k_1$, $k_2$, and $k_3$ are along the [100], [010], and [001] directions, respectively. The model is built on a face-centered cubic (fcc) lattice as the real material, and the distance between nearest neighbors on the fcc lattice is $\sqrt{2}a$, where $a$ is the length scale used in Eq.~\eqref{fourband}. This model Hamiltonian captures the band inversion of bulk states at $L$ points and the topological gapless surface states, as shown in Fig.~\ref{fig1}(b).  The energy difference between the surface Dirac points at the $\bar{\Gamma}$ and $\bar{M}_{1,2,3}$ points can be tuned by $t_3$.

The fcc lattice can be recast as the ABC stacking of a triangular lattice along the (111) direction, as shown in Fig~\ref{fig2}(a). The new coordinates are $x$, $y$, and $z$ along the [$11\bar{2}$], [$\bar{1}10$], and [111]  directions, respectively.
The onsite superconducting pairing potential term  is $\Delta(r, \theta)=\Delta_0 \tanh \left(r / \xi\right) e^{-i \theta}$, where $r=\sqrt{x^2+y^2}$ and $\xi$ is the coherence length. To reveal the $\mu$-tuned vortex phase transition, we plot the energies at $k_{z}=0$ and $k_{z}=\pi$ by taking the periodic (open) boundary condition along the $z$ ($x$ and $y$ directions) direction as a function of $\mu$ in Fig.~\ref{fig2}(b) and Fig.~\ref{fig2}(c), respectively. Here $k_z$ is in unit of $1/(2\sqrt{3}a)$, where $2\sqrt{3}a$ is the lattice period along the [111] direction.
A large energy gap at $k_{z}=0$ persists for different $\mu$. In contrast, with the evolution of $\mu$, the energy gap at $k_{z}=\pi$ closes and reopens at $\mu_c^{(1)}\approx -0.7$, $\mu_c^{(2)}\approx -0.58$, $\mu_c^{(3)}\approx 0.38$, and $\mu_c^{(4)}\approx 0.7$, resulting in a series of topological phase transitions. The presence of four different critical chemical potentials is consistent with the analysis based on the bulk effective Hamiltonian in Eq.~\eqref{HkDirac}. The gap closes at $k_z=\pi$ since the four $L$ points are all projected onto the momentum $k_z=\pi$ for the wire system along the (111) direction. In particular, the orange bands in Fig.~\ref{fig2}(c) with the gap closing at $\mu_c^{(1)}$ and $\mu_c^{(4)}$ are threefold degenerate, which implies that these bands are contributed by the $L_{1,2,3}$ valleys. In contrast, the green bands in Fig.~\ref{fig2}(c) with the gap closing at $\mu_c^{(2)}$ and $\mu_c^{(3)}$ do not have degeneracy and therefore are generated by the $L_0$ valley.

The four-band model has the cubic symmetry as the real materials, which hosts the three-fold rotational symmetry around the [111] direction ($z$), denoted as $C_{3}$, and mirror symmetry $M_{y}$ which flips $y$ to $-y$. The system respects the chiral symmetry $\mathcal{S}$ and rotational symmetry $\mathcal{C}_3$ when considering the superconducting vortex along the [111] direction. In Appendix~\ref{Appendixd}, we present the concrete form of  $\mathcal{S}$ and $\mathcal{C}_3$.
By the $\mathcal{S}$ and $\mathcal{C}_3$ symmetries, we calculate the winding numbers $W_{3}^{(0,1,2)}$ with the evolution of $\mu$, as shown in Fig.~\ref{fig2}(d). We numerically confirm that $W_{3}^{(0)}=W_{3}^{(1)}=W_{3}^{(2)}$ as required by Eq.~\eqref{Wequi}. The gap closing and reopening shown in Fig.~\ref{fig2}(c) leads to topological phase transitions that are characterized by the change of the winding number 
$W_{3}^{(0)}$. In the interval $[\mu_c^{(1)},\mu_c^{(2)}]$ and $[\mu_c^{(3)},\mu_c^{(4)}]$, $W_{3}^{(0)}=1$. In the interval $[\mu_c^{(2)},\mu_c^{(3)}]$, $W_{3}^{(0)}=2$. For the former case with $W_{3}^{(0)}=1$, there is only one robust MVEMs. For the latter case, both $\tilde{\gamma}_0$ and $\tilde{\gamma}_1$ obtained in Eq.~\eqref{gamma0123} from the surface effective model 
are robust and protected by the $\mathcal{S}$ symmetry. It is noted that the vortex phase transitions can also be captured by the pfaffian $Z_2$ topological invariant. The pfaffian topological invariant is defined by \cite{Kitaev2001}
\beqn
(-1)^{\nu}=\text{sign}[\text{pf}(H_{M}(k_{z}=0))\text{pf}(H_{M}(k_{z}=\pi))],
\eeqn
where $\text{pf}(A)$ denotes the pfaffian value of an anti-symmetric matrix $A$ and $H_{M}$ denotes the Hamiltonian of the whole system in the Majorana basis. The pfaffian topological invariant $\nu$ is simply the parity of $W_{3}^{(0)}$ \cite{Tewari2012}, as shown in Fig.~\ref{fig2}(e). By fixing $\mu=0$ and diagonlizing $H_{\text{BdG}}$ under the open boundary conditions, we obtain eight states near zero-energy, which are localized at the vortex core, as shown in Fig.~\ref{fig2}(f). This implies that there are four MVEMs on the (111) surface. However, only two of them are protected by the the chiral symmetry and the other two can be hybridized without symmetry breaking.

In-plane fields can be used to to detect and manipulate MVEMs \cite{Liu2024}.  When further applying an in-plane magnetic field perpendicular to
the [111] direction, 
the vortex axis in principle is no-longer along the [111] direction. For the in-plane field along a generic direction, the translational symmetry along the [111] direction, the chiral symmetry $\mathcal{S}$, and rotational symmetry  $\mathcal{C}_3$ are explicitly broken. The Majorana vortex phase characterized by $W_{3}^{(0)}=2$ can be trivialized. While the topological phase characterized by $W_{3}^{(0)}=1$ 
is essentially protected by the particle-hole symmetry.  As long as the energy gap of the system is not closed, unpaired MVEMs preserves.

\section{Discussion and summary }
\label{SecV1}
Breaking crystal symmetry in TCIs provides a versatile approach to engineering various topological phases \cite{Mandal2017,Schindler2018,Lau2019,Mogi2024,Krizman2024}. For instance,  Bi-doping of Pb$_{1-x}$Sn$_x$Se (111) epilayers can realize a strong TI phase in which the Dirac cone at $\bar{\Gamma}$ is gapped while the three Dirac cones at $\bar{M}_{1,2,3}$ remain intact \cite{Mandal2017}. Moreover, it is theoretically predicted that uniaxial strain can drive SnTe to be a higher-order TI hosting helical hinge states \cite{Schindler2018}. Weyl semimetal can also be realized by considering a lattice distortion along the [111] direction \cite{Lau2019}. Notably, all these topological phases can be used to realize MVEMs \cite{Fu2008,Yanzhongbo2020,Ghorashi2020}. Thus, an interesting direction is to explore the interplay between Majorana vortex phases and different topological band structures induced by symmetry breaking.

Zero-basis conductance peak is an important evidence for the observation of MZMs \cite{LawKT2009}. On the (111) surface, the number of robust MZMs can change from $2\rightarrow 1\rightarrow 0$ by tuning chemical potential. The phases hosting one and two robust MZMs are associated with zero-basis conductance peak $4e^2/h$ and $2e^2/h$, respectively, at zero
temperature. This implies that across the transition point, the zero basis conductance will have a quantized change, which can provide a potential evidence for MVEMs.

We emphasize that the topological phase transitions depend on the details of the system, including doping \cite{Mandal2017}, boundary cleavage \cite{liujunwei2013}, and strain effect \cite{Zhaolu2014}. In our study, the four-band model is used to capture the essential physics in SnTe. However, more realistic models and refined parameters are necessary for a precise determination of the Majorana vortex phase transitions in SnTe.

In summary, motivated by the recent experiment progress for the observation and manipulation of MVEMs in SnTe with proximity-induced superconductivity \cite{Liu2024}, we present a comprehensive study of the Majorana phases on different surfaces. Our study includes topological classification, effective model analysis,  and lattice model calculations. We find that a series of vortex phase transitions can occur for the [111]-direction vortex tuned by the chemical potential.
Our work could provide theoretical guidance for future experimental studies of Majorana vortex phases on the (111) surface, which can be a rich platform.
The crystal symmetry-protected Majorana vortex phases could also be realized in other TCI systems with proximity-induced superconductivity, for example, the twofold rotational symmetry-protected TCI of bismuth \cite{ChuangHanHsu2019}.

\section{Acknowledgments}
This work is supported by National Key Research and Development Program
of China (Grant No. 2022YFA1402401).

\appendix

\section{Relations between two sets of winding numbers}
\label{windingnumberrelations}
The winding number for the BdG Hamiltonian $H_{\text{BdG}}$ is defined by 
\beqn
W=\frac{1}{4\pi i}\int_{0}^{2\pi}dk_z\text{Tr}[\mathcal{S}H_{\text{BdG}}^{-1}\partial_{k_z}H_{\text{BdG}}].
\eeqn
Because $H_{\text{BdG}}$ respects the chiral symmetry $\mathcal{S}_n^{(p)}=\mathcal{C}_n^{n}\mathcal{S}$,  we can define $n$ winding numbers $W_{n}^{(p)}$ by replacing $\mathcal{S}$ with $\mathcal{S}_n^{(p)}$, where $p=0,\cdots,n-1$ and $\mathcal{S}_n^{(0)}=\mathcal{S}$. Because $[\mathcal{C}_n, H_{\text{BdG}}]=0$, $H_{\mathrm{BdG}}$
 can be block-diagonalized into sectors spanned by eigenvectors of $\mathcal{C}_{n}$, namely,
\beqn
 H_{\mathrm{BdG}}=H_{\mathrm{BdG}}^{(0)} \oplus H_{\mathrm{BdG}}^{(1)} \oplus \ldots \oplus H_{\mathrm{BdG}}^{(n-1)},
\eeqn
where $ H_{\mathrm{BdG}}^{(j)}$ represents the Hamiltonian in the subsector with eigenvalue $e^{i 2 \pi j/ n}$ of $\mathcal{C}_n$, for $j=0,\cdots,n-1$. Because of $\mathcal{C}_n\mathcal{M}_{\mathcal{T}}=\mathcal{M}_{\mathcal{T}}\mathcal{C}_n^{-1}$ and $[\mathcal{C}_n,P]=0$, we have $\mathcal{C}_n\mathcal{S}_{n}^{(p)}=\mathcal{S}_{n}^{(p)}\mathcal{C}_n^{-1}$, where $\mathcal{S}_{n}^{(p)}=\mathcal{C}_n^{p}\mathcal{M}_{\mathcal{T}}P=\mathcal{C}_n^{p}M_yTP$. Here $M_y$, $T$, and  $P$ are the mirror symmetry, the time-reversal symmetry, and particle-hole symmetry, respectively. Therefore, only for the block associated with real eigenvalue $e^{i2\pi j^{\prime}/n}$, $H_{\mathrm{BdG}}^{(j^{\prime})}$ respects the chiral symmetry $\mathcal{S}_{n}^{(p)}$, and the blocks associated with complex eigenvalues $e^{i 2 \pi j/ n}$ and $e^{-i 2 \pi j/ n}$ are related by the chiral symmetry $\mathcal{S}_n^{(p)}$. 
Therefore, only the block associated with real $e^{i2\pi j^{\prime}/n}$ has a contribution to $W_{n}^{(p)}$. For the Hamiltonian $H_{\mathrm{BdG}}^{(j^{\prime})}$, we can define the winding number
\beqn
w_{j^{\prime}}= \frac{1}{4\pi i}\int_{0}^{2\pi}dk_z\text{Tr}[\mathcal{S}(H_{\text{BdG}}^{(j^{\prime}}))^{-1}\partial_{k_z}H_{\text{BdG}}^{(j^{\prime})}].
\eeqn
With this definition, we have $W_{n}^{(0)}=\sum_{j^{\prime}}w_{j^{\prime}}$. The definition of $W_{n}^{(1)}$ is 
\beqn
W_{n}^{(1)}=\frac{1}{4\pi i}\int_{0}^{2\pi}dk_z\text{Tr}[\mathcal{S}_{n}^{(1)}H_{\text{BdG}}^{-1}\partial_{k_z}H_{\text{BdG}}]\nonumber\\
=\frac{1}{4\pi i}\int_{0}^{2\pi}dk_z\text{Tr}[\mathcal{S}\mathcal{C}_nH_{\text{BdG}}^{-1}\partial_{k_z}H_{\text{BdG}}].
\eeqn
For $n=4$, $\mathcal{C}_4$ has the eigenvalue of $1$ and $-1$, respectively, for the blocks of $j^{\prime}=0$ and $2$. Thus, we have
\begin{widetext}
\beqn
W_{4}^{(1)}&=&\frac{1}{4\pi i}\int_{0}^{2\pi}dk_z\text{Tr}[\mathcal{C}_4 \mathcal{S}H_{\text{BdG}}^{-1}\partial_{k_z}H_{\text{BdG}}]\nonumber\\
&=&\frac{1}{4\pi i}\int_{0}^{2\pi}dk_z\text{Tr}[\mathcal{S}(H_{\text{BdG}}^{(0)})^{-1}\partial_{k_z}H_{\text{BdG}}^{(0)}]
-\text{Tr}[\mathcal{S}(H_{\text{BdG}}^{(2)})^{-1}\partial_{k_z}H_{\text{BdG}}^{(2)}]\nonumber\\
&=&w_{0}-w_{2}.
\eeqn
\end{widetext}
Similarly, we have $W_{2}^{(1)}=w_{0}-w_{1}$.

\section{Majorana vortex end modes on the (001) surface }
\label{appendixa}
\begin{figure}[b]
\includegraphics[width=3.3in]{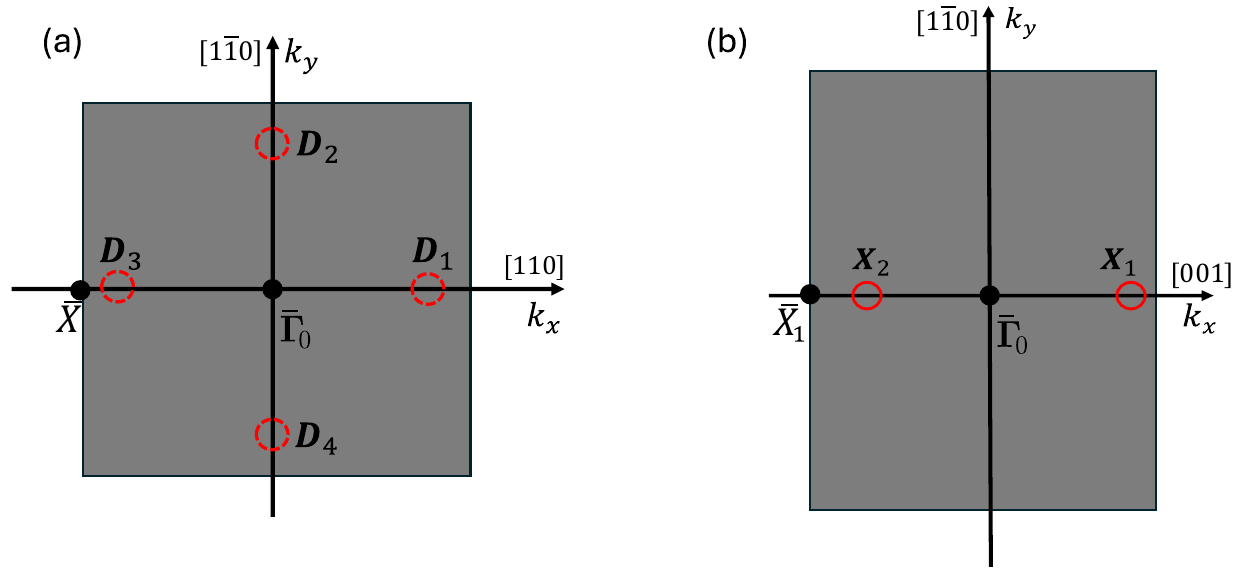}
\caption{(a) and (b) The surface Brillouin zone of the (001) and (110) surfaces and schematic illustration of the energy contour of the surface states around Dirac cones. }
\label{fig3}
\end{figure}

SnTe hosts four surface Dirac cones on the (001) surface, as schematically illustrated in Fig.~\ref{fig3}(a).  Two Dirac cones along the line $k_y=0$ [i.e., [$110$] direction], are denoted by $\bm{D}_{1,3}$, and the other two along the line  $k_x=0$ [i.e., $[1\bar{1}0]$ direction], are denoted by $\bm{D}_{2,4}$. The Hamiltonian of the
$\bm{D}_i$ surface Dirac cone is 
\beqn
\hat{h}_{i}=\sum_{|\bm{q}|<\Lambda, s, s^{\prime}=\uparrow, \downarrow} h_{i}^{s s^{\prime}}(\bm{q}) D_{i s}^{\dagger}(\bm{q}) D_{i s^{\prime}}(\bm{q}),
\eeqn
where $D_{is}(\bm q)$ is the annihilation operator at $\bm{k}=\bm{D}_{i}+\bm{q}$ with pesudospin $s$, for $i=1,2,3,4$. The form of $h_{1,2,3,4}$ can be derived by choosing the representation of the $C_{4v}$ and $T$ symmetries and we follow the chosen representation in Ref.~\cite{Fang2014}. In particular, $h_1$ is fixed by choosing the representation of the little group at $\bm{D}_1$ to be $M_y=is_y$ and $C_2T=s_xK$.  $h_{2,3,4}$ are fixed by choosing the gauges such that ${D}_{2,3,4}(C_4\bm q)=C_4{D}_{1,2,3}(\bm q)C_4^{-1}$, where $C_4(q_x,q_y)\rightarrow (-q_y,q_x)$ and  the index $s$ is implicit for convenience. The chosen representations and gauges lead to \cite{Fang2014}
\begin{equation}
\begin{aligned}
&h_1(\bm q)=v(q_xs_y-q_ys_x),\\
&h_2(\bm q)=h_1(C_4^{-1}\bm q)=v(q_ys_y+q_xs_x),\\
&h_3(\bm q)=h_1(-C_2\bm q)=-v(q_xs_y-q_ys_x),\\
&h_4(\bm q)=h_1(C_4\bm q)=-v(q_ys_y+q_xs_x).
\end{aligned}
\end{equation}
Here we assume that the Fermi velocity is isotropic. In the basis of $\{D_{1}(\bm q), D_{3}(\bm q),D_{2}(\bm q),D_{4}(\bm q)\}$, the total Hamiltonian of describing the four surface Dirac cones can be written as
\beqn
h=&(&\rho_0+\rho_z)/2(q_x\eta_zs_y-q_y\eta_zs_x)\nonumber\\
&+&(\rho_0-\rho_z)/2(q_y\eta_zs_y+q_x\eta_zs_x),
\eeqn
where Pauli matrices $\rho$ and $\eta$ act on the subspace expanded by the valleys related by a four-fold rotation and inversion, respectively.
$h$ respects the mirror, four-fold rotation, and time-inversion symmetries
\begin{equation}
\begin{aligned}
&M_{1\bar{1}0}^{-1}H(q_x,q_y)M_{1\bar{1}0}=H(M_{1\bar{1}0}\bm q)=H(q_x,-q_y),\\
&M_{110}^{-1}H(q_x,q_y)M_{110}=H(M_{110}\bm q)=H(-q_x,q_y),\\
&C_4^{-1}H(q_x,q_y)C_4=H(C_4\bm q)=H(-q_y,q_x),\\
&T^{-1}H(q_x,q_y)T=H(-q_x,-q_y).
\end{aligned}
\end{equation}
where
\begin{equation}
\begin{aligned}
&M_{1\bar{1}0}=i(\rho_0+\rho_z)/2\eta_zs_y-i(\rho_0-\rho_z)/2\eta_xs_y, \\
&M_{110}=-i(\rho_0+\rho_z)/2\eta_xs_y-i(\rho_0-\rho_z)/2\eta_zs_y, \\
&C_4=(\rho_x+i\rho_y)/2\eta_0s_0+i(\rho_x-i\rho_y)/2\eta_ys_0,\\
&T=-i\rho_0\eta_ys_xK.
\end{aligned}
\end{equation}
These symmetries satisfy the the relations $[T,C_4]=0$, $\{M_{110},C_4^2\}=0$, $\{M_{1\bar{1}0},C_4^2\}=0$, and $M_{110}=C_4^2M_{1\bar{1}0}$.

The superconducting pairing for a conventional $s$-wave superconductor is formed between $D_{1\uparrow}(\bm q)$ [$D_{2\uparrow}(\bm q)$] and $D_{3\downarrow}(-\bm q)$ [$D_{4\downarrow}(-\bm q)$]. 
In the basis of $\{D_{1}(\bm q),D_{3}(\bm q),D_{1}^{\dagger}(-\bm q),D_{3}^{\dagger}(-\bm q)\}^{T}$,  the BdG Hamiltonian with superconducting vortex can be written as
\beqn
H_{\text{BdG}}=&v&q_x\tau_z\eta_zs_y-vq_y\tau_0\eta_zs_x\nonumber\\
&+&\Delta_0\tanh (r/\xi)(\frac{x}{r}\tau_y\eta_ys_x-\frac{y}{r}\tau_x\eta_ys_x).
\eeqn
Without loss of generality, we assume that $v$ and $\Delta_0$ have the identical sign. $H_{\text{BdG}}$ respects the chiral symmetry $\mathcal{S}=M_{1\bar{1}0}PT=-\tau_x\eta_xs_z$, where $M_{1\bar{1}0}=i\tau_0\eta_zs_y$ and $P=\tau_x K$. $H_{\text{BdG}}$ is block-diagonal and can be further written as
\begin{widetext}
\begin{equation}
\begin{aligned}
&H_{\text{BdG}}=h_{+}\bigoplus h_{-}, \quad h_{-}=-h_{+},\nonumber\\
&h_{+}=\left(\begin{matrix}
0 & -ve^{-i\theta}(\partial_r-i/r\partial_{\theta}) & 0 & -\Delta( r)e^{-i\theta}\\
ve^{i\theta}(\partial_r+i/r\partial_{\theta}) & 0 & -\Delta( r)e^{-i\theta} & 0\\
0 & -\Delta( r)e^{i\theta} & 0 & -ve^{i\theta}(\partial_r+i/r\partial_{\theta})\\
-\Delta( r)e^{i\theta} & 0 & ve^{-i\theta}(\partial_r-i/r\partial_{\theta})& 0
\end{matrix}\right),
\end{aligned}
\end{equation}   
\end{widetext}
where $\Delta( r)=\Delta_0\tanh r/\xi$. A zero-energy solution can be derived from $h_{+}$ and $h_{-}$, respectively, 
\begin{equation}
\begin{aligned}
&\kappa_1=(D_{1\downarrow}+D_{3\downarrow}^{\dagger})e^{-\int_{0}^{r}\Delta( r^{\prime})/vdr^{\prime}},\\
&\kappa_3=\kappa_1^{\dagger}=(D_{3\downarrow}+D_{1\downarrow}^{\dagger})e^{-\int_{0}^{r}\Delta( r^{\prime})/vdr^{\prime}}.
\end{aligned}
\end{equation}
From the superposition of $\kappa_1$ and $\kappa_3$, we can obtain two MZMs solution 
\begin{equation}
\begin{aligned}
&\gamma_1=\kappa_1+\kappa_3=(D_{1\downarrow}+D_{3\downarrow}+h.c.)e^{-\int_{0}^{r}\Delta(r^{\prime})/vdr^{\prime}},\\
&\gamma_3=i(\kappa_3-\kappa_1)=[i(D_{3\downarrow}-D_{1\downarrow})+h.c.]e^{-\int_{0}^{r}\Delta( r^{\prime})/vdr^{\prime}}.
\end{aligned}
\end{equation}
The spinor part of MZMs $\gamma_{1}$ and $\gamma_{3}$, respectively, is 
\begin{equation}
\begin{aligned}
&\psi_1=\frac{1}{4}[0,1,0,1,0,1,0,1]^{T},\\
&\psi_2=\frac{1}{4}[0,-i,0,i,0,i,0,-i]^{T}.
\end{aligned}
\end{equation}
It can be checked that $\mathcal{S}\psi_{1,2}=\psi_{1,2}$, which implies that the MZMs $\gamma_{1,3}$ is invariant under the chiral symmetry $\mathcal{S}$. Similarly, two MZMs can be obtained when considering the superconducting pairing between $D_{2\uparrow}(\bm q)$ and $D_{4\downarrow}(-\bm q)$, which read as
\begin{equation}
\begin{aligned}
&\gamma_2=(e^{i\pi/4}D_{2\downarrow}+e^{i\pi/4}{D}_{4\downarrow}+h.c.)e^{-\int_{0}^{r}\Delta( r^{\prime})/vdr^{\prime}},\\
&\gamma_4=(e^{i3\pi/4}D_{4\downarrow}-e^{i3\pi/4}D_{2\downarrow}+h.c.)e^{-\int_{0}^{r}\Delta( r^{\prime})/vdr^{\prime}}.
\end{aligned}
\end{equation}
 It can be shown that the MZMs $\gamma_{2,4}$ transform into $\gamma_{4,2}$ under the chiral symmetry $\mathcal{S}$. Thus, in the zero-energy subspace expanded by $\gamma_{1,2,3,4}$, $\mathcal{S}$ is represented by 
\beqn
\mathcal{S}=\left(\begin{matrix}
1 & 0 & 0 & 0\\
0 & 0 & 0 & 1\\
0 & 0 & 1 & 0\\
0 & 1 & 0 & 0
\end{matrix}\right).
\eeqn
Incorporating the four Dirac cones, the BdG Hamiltonian of the whole system hosts the four-fold rotational symmetry $\mathcal{C}_4$, which is defined by 
\beqn
&&\mathcal{C}_4D_{1,2,3}\mathcal{C}_4^{-1}=e^{i\pi/4}D_{2,3,4},\nonumber\\ &&\mathcal{C}_4D_4\mathcal{C}_4^{-1}=-e^{i\pi/4}D_1,
\eeqn
where $\mathcal{C}_4^4=1$. Thus,  in the zero-energy subspace expanded by $\gamma_{1,2,3,4}$, $\mathcal{C}_4$ is represented by 
\beqn
\mathcal{C}_4=\left(\begin{matrix}
0 & 1 & 0 & 0\\
0 & 0 & 1 & 0\\
0 & 0 & 0 & 1\\
1 & 0 & 0 & 0
\end{matrix}\right).
\eeqn
Since $[\mathcal{C}_4, H_{\text{BdG}}]=0$ and $\{\mathcal{S},H_{\text{BdG}}\}=0$, the operators $\mathcal{C}_4\mathcal{S}$, $\mathcal{C}_4^{2}\mathcal{S}$, and $\mathcal{C}_4^{3}\mathcal{S}$ are also the chiral symmetry of $H_{\text{BdG}}$ and we have 
\beqn
&&\text{Tr}(\mathcal{S})=\text{Tr}(\mathcal{C}_4^{2}\mathcal{S})=2, \nonumber\\
&&\text{Tr}(\mathcal{C}_4\mathcal{S})=\text{Tr}(\mathcal{C}_4^{3}\mathcal{S})=0.
\eeqn
This implies that two out of the four MZMs are protected by the chiral symmetries $\mathcal{S}$ and $\mathcal{C}_4^{2}\mathcal{S}$, which are generated by the $\mathcal{M_T}$ symmetry with the mirror along the $(1\bar{1}0)$ and $(110)$ directions, respectively. While the chiral symmetries associated with the (100) and (010) direction mirror symmetries can not protect MZMs.

By diagonlizing $\mathcal{C}_4$, we can obtain four MZMs with fixed angular momentum, which are
\begin{equation}
\begin{aligned}
&\tilde{\gamma}_1=(\gamma_1+\gamma_2+\gamma_3+\gamma_4)/2,\\
&\tilde{\gamma}_2=(\gamma_1-\gamma_2+\gamma_3-\gamma_4)/2,\\
&\tilde{\gamma}_3=(i\gamma_1+\gamma_2-i\gamma_3-\gamma_4)/2,\\
&\tilde{\gamma}_4=(-i\gamma_1+\gamma_2+i\gamma_3-\gamma_4)/2,
\end{aligned}
\end{equation}
where $\tilde{\gamma}_{1}$, $\tilde{\gamma}_{2}$, $\tilde{\gamma}_{3}$, and $\tilde{\gamma}_{4}$ have the angular momentum $J_z=0$, $J_z=2$, $J_z=1$, and $J_z=3$, respectively.  It can be shown that  
\begin{equation}
\begin{aligned}
&\mathcal{S}\tilde{\gamma}_{1,2}\mathcal{S}^{-1}=\tilde{\gamma}_{1,2}, \\
&\mathcal{S}\tilde{\gamma}_{3,4}\mathcal{S}^{-1}=-\tilde{\gamma}_{4,3}, \\
&\mathcal{C}_4^2\mathcal{S}\tilde{\gamma}_{1,2}(\mathcal{C}_4^2\mathcal{S})^{-1}=\tilde{\gamma}_{1,2},\\
&\mathcal{C}_4^2\mathcal{S}\tilde{\gamma}_{3,4}(\mathcal{C}_4^2\mathcal{S})^{-1}=\tilde{\gamma}_{4,3}.
\end{aligned}
\end{equation}
This implies that the MZMs $\tilde{\gamma}_{1,2}$ are protected by  $\mathcal{S}$ and  $\mathcal{C}_4^2\mathcal{S}$ symmetries while $\tilde{\gamma}_{3,4}$ are not. Because possessing different angular momentum, the MZMs $\tilde{\gamma}_{1.2}$  are also protected by the  rotational symmetry. Thus,  two out of the four MZMs are protected by both the chiral and rotational symmetries.

\begin{figure}[t]
\centering
\includegraphics[width=3.3in]{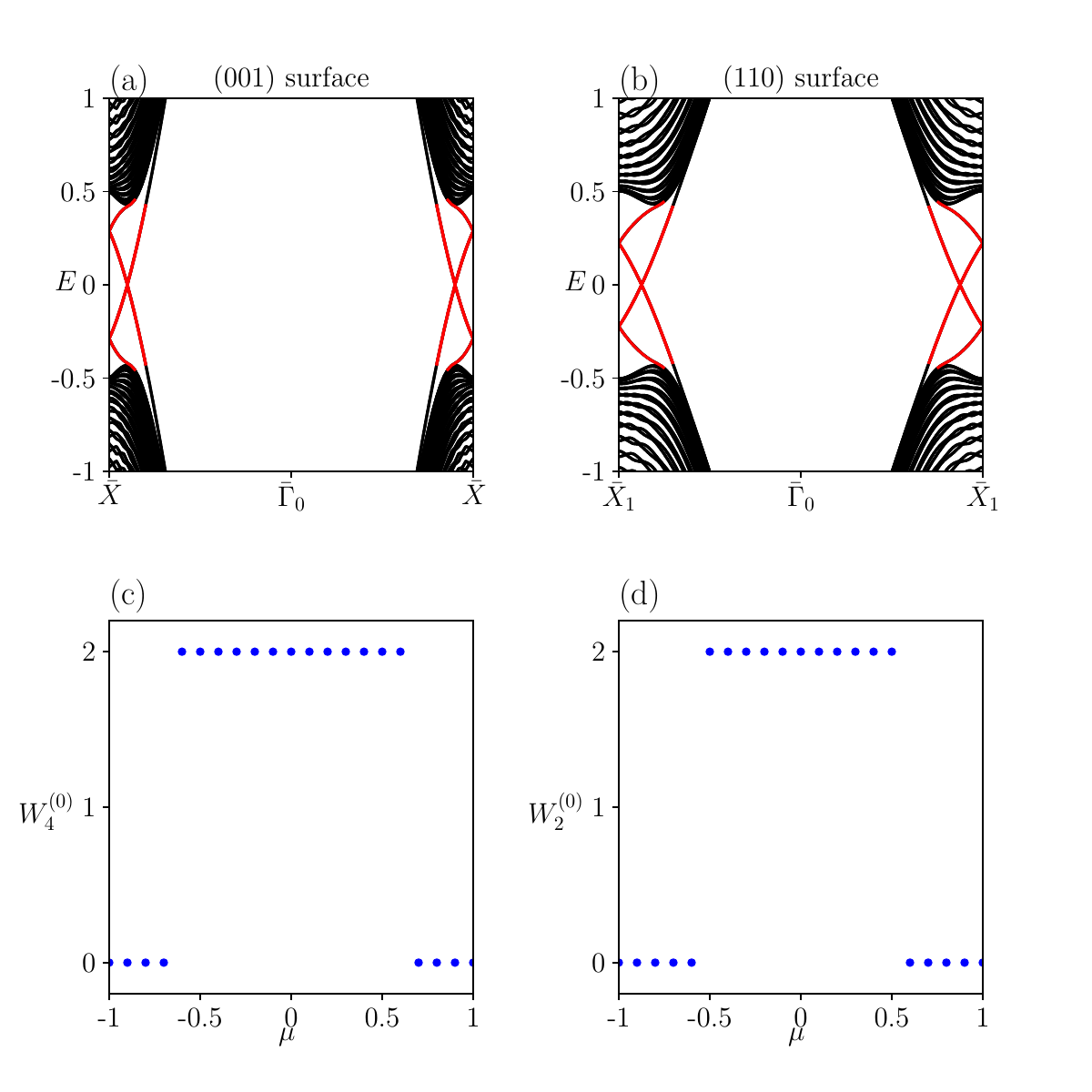}
\caption{(a) and (b) The energy spectra of $H_{\text{BdG}}$ with a slab geometry on the (001) and (110) surfaces, respectively. (c) and (d) Winding numbers $W_{4}^{(0)}$ and $W_{2}^{(0)}$ for the [001] and [110] direction vortex, respectively, as functions of $\mu$. The model parameters are taken as $m=2.5$, $t_1=-1$, $t_2=1$, $t_3=0$, $\Delta_0=0.5$, and $\xi=2$. }
\label{fig4}
\end{figure}

\section{Symmetries analysis of four-band model and winding number calculation}
\label{Appendixd}
The model Hamiltonian of the four-band model  is 
\begin{equation}
\begin{aligned}
h(\mathbf{k}) & =[m-t_1(\cos 2 k_1+\cos  2k_2+\cos 2k_3)] \sigma_{z}s_0 \\
&+t_2[\sin k_1\left(\cos k_2+\cos k_3\right) \sigma_{x }s_x \\
&+\sin k_2(\cos k_1+\cos k_3) \sigma_{x}s_y\\
&+\sin k_3(\cos k_1+\cos k_2) \sigma_{x}s_z]\\
&+t_3[\cos k_1(\cos k_2+\cos k_3)+\cos k_2\cos k_3]\sigma_0s_0.
\end{aligned}
\end{equation}
$h$ is defined on a face-centered cubic lattice and has the cubic symmetry. The generators are the four-fold rotation around the $(001)$ direction $C_{4 z}=\sigma_0 \otimes e^{-i \pi s_z / 4}$, the four-fold rotation around the $(100)$ direction $C_{4 x}=\sigma_0 \otimes e^{-i\pi s_x/ 4}$, and the inversion $I=\sigma_{z}s_0$. When considering the superconducting vortex along different directions, the BdG Hamiltonian can be generally written as
\beqn
H_{\mathrm{BdG}}(x,y,k_{z})=\left(\begin{array}{cc}
h\left(x,y,k_{z}\right)-\mu & \Delta(r, \theta) i s_y \\
-\Delta^*(r, \theta) i s_y & -h^T\left(x,y,-{k}_{z}\right)+\mu
\end{array}\right),\nonumber\\
\label{ha2}
\eeqn
where $\Delta(r, \theta)=\Delta_0 \tanh \left(r / \xi\right) e^{-i \theta}$ with $r=\sqrt{x^2+y^2}$ and $\theta$ being the polar angle.
For the vortex in the [001] direction, we align the unit vectors as follows: $\boldsymbol{e}_x$ along the $[\bar{1} \bar{1} 0]$ direction, $\boldsymbol{e}_y$ along the $[1 \bar{1} 0]$ direction, and $\boldsymbol{e}_z$ along the [001] direction. For the vortex in the [110] direction, we choose $\boldsymbol{e}_x$ along the [001] direction, $\boldsymbol{e}_y$ along the [$1 \bar{1} 0$] direction, and $\boldsymbol{e}_z$ along the [110] direction. Finally, for the vortex in the [111] direction, $\boldsymbol{e}_x$ is aligned along the $[11 \bar{2}]$ direction, $\boldsymbol{e}_y$ along the $[1 \bar{1} 0]$ direction, and $\boldsymbol{e}_z$ along the [111] direction.  Here, $\bm{e}_{x,y,z}$ are the coordinate basis of the $x-y-z$ 
coordinate frame.

For the [001] direction vortex, the $\mathcal{C}_4$ and $\mathcal{S}$ symmetries of $H_{\text{BdG}}$ are given by
\beqn
&&\mathcal{C}_4=R_4(x,y)\otimes \tilde{\mathcal{C}}_4, \nonumber\\
&&\quad\mathcal{S}=m_{1\bar{1}0}(x,y)\mathcal{M}_{1\bar{1}0}TP,\nonumber\\
&&\tilde{\mathcal{C}}_4=\left(\begin{array}{cc}
e^{-i\pi/4 }C_{4z} & 0 \\
0 & e^{i\pi/4 }(C_{4z})^{\ast}\\
\end{array}\right), \nonumber\\
&&\mathcal{M}_{1\bar{1}0}=\left(\begin{array}{cc}
{M}_{1\bar{1}0} & 0 \\
0 & ({M}_{1\bar{1}0})^{*}
\end{array}\right),
\eeqn
where $R_4$ and $m_{1\bar{1}0}$, respectively, act on the spatial coordinate as $R_4(x,y)\rightarrow (-y,x)$ and $m_{1\bar{1}0}(x,y)\rightarrow (-y,x)$. $T$, $P$, and $M_{1\bar{1}0}$ are given by $T=is_yK$, $P=\tau_xK$, and $M_{1\bar{1}0}=i\sigma_zs_x e^{-i \pi s_z / 4}$. For the (110) direction vortex, the $\mathcal{C}_2$ symmetry of $H_{\text{BdG}}$ is
\begin{equation}
\begin{aligned}
&\mathcal{C}_2=R_2(x,y)\otimes \tilde{\mathcal{C}}_2,\\
&\tilde{\mathcal{C}}_2=\left(\begin{array}{cc}
e^{-i\pi/2 }C_{2z} & 0 \\
0 & e^{i\pi/2 }(C_{2z})^{\ast}\\
\end{array}\right), 
\end{aligned}
\end{equation}
where $R_2$ acts on the spatial coordinate as $R_2(x,y)=(-x,-y)$,  and $C_{2z}=e^{-i\pi/2(s_x+s_y)/\sqrt 2}$.
For the (111)-direction vortex, the $\mathcal{C}_3$ symmetry of $H_{\text{BdG}}$ is
\begin{equation}
\begin{aligned}
&\mathcal{C}_3=R_3(x,y)\otimes \tilde{\mathcal{C}}_3,\\
&\tilde{\mathcal{C}}_3=\left(\begin{array}{cc}
e^{-i\pi/3 }C_{3z} & 0 \\
0 & e^{i\pi/3 }(C_{3z})^{\ast}\\
\end{array}\right), 
\end{aligned}
\end{equation}
where $R_3$ acts on  the spatial coordinate as $R_3(x,y)\rightarrow(-x/2-\sqrt{3}y/2, \sqrt{3}x/2-y/2)$ and $C_{3z}=e^{-i\pi (s_x+s_y+s_z)/3\sqrt{3}}$.
 For the [111] and [110] directions vortex, $H_{\text{BdG}}$ also respects the chiral symmetry $\mathcal{S}$.

In Fig.~\ref{fig4}(a) and \ref{fig4}(b), we present the surface states for 
 a slab geometry on the (001) and (110) surfaces. In Fig.~\ref{fig4}(c) and \ref{fig4}(d), we calculate the winding numbers $W_{4}^{(0)}$ and $W_{2}^{(0)}$ for the [001] and [110] direction vortex as functions of $\mu$.

\section{Majorana vortex end modes on the (110) surface}
\label{Appendixb}
SnTe hosts two surface Dirac cones on the (110) surface \cite{liujunwei2013}, denoted by $\bm{X}_{1}$ and $\bm{X}_{2}$, as schematically
illustrated in Fig.~\ref{fig3}(b). Similar to the Dirac cones $\bm{D}_{1,3}$ on the (001) surface, Dirac points at $\bm{X}_{1,2}$ are invariant under the mirror $M_{1\bar{1}0}$ and $C_{2}T$ symmetries. Therefore, the symmetries analysis of 
the MZMs contributed by Dirac cones $\bm{D}_{1,3}$ can be directly applied to the MZMs contributed by the Dirac cones $\bm{X}_{1,2}$. It can be shown that there are two MZMs on the (110) surface, denoted by $\gamma_1$ and $\gamma_2$. In the zero-energy subspace, the chiral symmetry $\mathcal{S}$ and two-fold rotational symmetry $\mathcal{C}_2$ can be represented by 
\beqn
\mathcal{S}=\left(\begin{matrix}
1 & 0 \\
0 & 1 \\
\end{matrix}\right),\quad\mathcal{C}_2=\left(\begin{matrix}
0 & 1 \\
1 & 0 \\
\end{matrix}\right),
\eeqn
which yields $\text{Tr}(\mathcal{S})=2$ and $\text{Tr}(\mathcal{C}_2\mathcal{S})=0$. The two MZMs $\tilde{\gamma}_1=(\gamma_1+\gamma_2)/\sqrt2$ and $\tilde{\gamma}_2=(\gamma_1-\gamma_2)/\sqrt 2$ are the common eigenstates of $\mathcal{S}$ and $\mathcal{C}_2$ symmetries, with the identical and different eigenvalues, respectively. Therefore, the two MZMs $\tilde{\gamma}_{1,2}$ are protected by both  $\mathcal{S}$ and $\mathcal{C}_2$ symmetries.
 
\section{Majorana vortex modes on the (111) surface}
\label{Appendixc}
SnTe hosts four Dirac cones on the (111) surface which are located at the $\bar{\Gamma}$ and $\bar{M}_{1,2,3}$ points, respectively. The surface Hamiltonian of describing the Dirac cone centered at $\bar{\Gamma}$ and $\bar{M}_{1,2,3}$ can be written as
\beqn
\hat{h}_{i=0,1,2,3}=\sum_{|\bm{q}|<\Lambda, s, s^{\prime}=\uparrow, \downarrow} h_{i}^{s s^{\prime}}(\bm{q}) f_{i s}^{\dagger}(\bm{q}) f_{i s^{\prime}}(\bm{q}),
\eeqn
where $f_{0 s}(\bm{q})$ and $f_{j s}(\bm{q})$ is the annihilation operator at $\bm k=\bar{\Gamma}+\bm{q}$ and $\bm k=\bar{M}_{j}+\bm{q}$, respectively, for $j=1,2,3$.   $h_0$ is fixed by choosing the representation of the little group at $\bar{\Gamma}$ to be $T=is_yK$,  $M_y=is_y$, and  $C_3=e^{i\pi s_z/3}$, which generate the constraints
\begin{equation}
\begin{aligned}
&T^{-1}h_0(\bm q)T=h_0(-\bm q),\\
&M_y^{-1}h_0(q_x,q_y)M_y=h_0(q_x,-q_y),\\
&C_3^{-1}h_0(\bm q)C_3=h_0(C_3\bm q),
\end{aligned}
\end{equation}
with  $C_3(q_x,q_y)\rightarrow (-q_x/2-\sqrt{3}q_y/2, \sqrt{3}q_x/2-q_y/2)$. Thus, we have $h_0(q)=v_1(q_xs_y-q_ys_x)$. The little group at $\bar{M}_1$ is generated by $T$ and $M_y$ symmetries. From the chosen representation,  we can derive $h_1=v_2(q_xs_y-q_ys_x)$. Here we assume isotropic Fermi velocity for the $\bar{M}_1$ valley. Using the $C_3$ symmetry, we can fix the gauges for Dirac cones centered at $\bar{M}_{2,3}$ as $C_3f_{1,2}(\bm q)C_3^{-1}=f_{2,3}(C_3\bm q)$.
Thus, we have 
\begin{equation}
\begin{aligned}
h_2(\bm q)&=h_1(C_3^{-1}\bm q)\\
&=(-q_x/2+\sqrt{3}q_y/2)s_y+(\sqrt{3}q_x/2+q_y/2)s_x,\\
h_3(\bm q)&=h_1(C_3\bm q)\\
&=(-q_x/2-\sqrt{3}q_y/2)s_y-(\sqrt{3}q_x/2-q_y/2)s_x.
\end{aligned}
\end{equation}
The mirror symmetry $M_y$ acts on the operators $f_{0,1,2,3}$ as
\begin{equation}
\begin{aligned}
&M_yf_{0,1}M_y^{-1}=is_yf_{0,1}(M_1q),\\
&M_yf_2M_y^{-1}=M_1C_3f_1C_3^{-1}M_1^{-1}=C_3^{-1}M_yf_1M_y^{-1}C_3=-is_yf_3,\\
&M_yf_3M_y^{-1}=M_1C_3f_2C_3^{-1}M_1^{-1}=C_3^{-1}M_yf_2M_y^{-1}C_3=-is_yf_2.
\end{aligned}
\end{equation}
where the relation $M_yC_3=C_3^{-1}M_y$ is used. The time-reversal symmetry $T$ acts on the operators $f_{0,1,2,3}$ as
\begin{equation}
\begin{aligned}
&Tf_{0,1}T^{-1}=is_yf_{0,1},\\
&Tf_2T^{-1}=TC_3f_1C_3^{-1}T^{-1}=C_3Tf_1T^{-1}C_3^{-1}=is_yf_2,\\
&Tf_3T^{-1}=TC_3f_2C_3^{-1}T^{-1}=C_3Tf_2T^{-1}C_3^{-1}=is_yf_3.\\
\end{aligned}
\end{equation}
where the relation $[T,C_3]=0$ is used.

When considering the superconducting vortex, the BdG Hamiltonian of describing the Dirac cone centered at $\bar{\Gamma}$  can be written as 
\begin{equation}
\begin{aligned}
H_{\text{BdG}}&=v_1(q_x\tau_zs_y-q_y\tau_0s_x)\nonumber\\
&+\Delta_0\tanh(r/\xi)(\frac{x}{r}\tau_ys_y-\frac{y}{r}\tau_xs_y).
\end{aligned}
\end{equation}
Here $H_{\text{BdG}}$ respects the particle-hole symmetry  $P=\tau_xK$ and three-fold rotational symmetry $\mathcal{C}_3=e^{i\pi/3}(\tau_0+\tau_z)/2e^{i\pi s_z/3}+e^{-i\pi/3}(\tau_0-\tau_z)/2e^{-i\pi s_z/3}$, where $\mathcal{C}_3^{-1}h_0(\bm r)\mathcal{C}_3=h_0(C_3\bm r)$ and $\mathcal{C}_3^3=1$. Moreover, $H_{\text{BdG}}$ respects the magnetic mirror symmetry $\mathcal{M_T}=K$, which leads to the chiral symmetry $\mathcal{S}=\mathcal{M_T}P=\tau_x$ ($y\rightarrow -y$).  $H_{\text{BdG}}$ can be further written as 
\begin{widetext}
\begin{equation}
\begin{aligned}
H_{\text{BdG}}=\left(\begin{array}{cccc}
0 & -v_1 e^{-i \theta}\left(\partial_r-\frac{i \partial_\theta}{r}\right) & 0 & -\Delta(r) e^{-i \theta} \\
v_1 e^{i \theta}\left(\partial_r+\frac{i \partial_\theta}{r}\right) & 0 & \Delta(r) e^{-i \theta} & 0 \\
0 & \Delta(r) e^{i \theta} & 0 & v_1 e^{i \theta}\left(\partial_r+\frac{i \partial_\theta}{r}\right) \\
-\Delta(r) e^{i \theta} & 0 & -v_1 e^{-i \theta}\left(\partial_r-\frac{i \partial_\theta}{r}\right) & 0
\end{array}\right),
\end{aligned}
\end{equation}    
\end{widetext}
where $\Delta( r)=\Delta_0\tanh r/\xi$. The MZM solution of $H_{\text{BdG}}$ is 
\begin{equation}
\begin{aligned}
&\gamma_0=(f_{0\downarrow}+f_{0\downarrow}^{\dagger})e^{-\int_{0}^{r}\Delta( r^{\prime})/v_1dr^{\prime}},
\end{aligned}
\end{equation}
where we assume that $\Delta_0$ and $v_1$ share the identical sign. It can be checked that $\mathcal{S}\gamma_0\mathcal{S}^{-1}=\gamma_0$ and $\mathcal{C}_3\gamma_{0}\mathcal{C}_3^{-1}=\gamma_{0}$. Similarly, the Dirac cone-centered at $\bar{M}_1$ with superconducting vortex gives rise to the MZM 
\begin{equation}
\begin{aligned}
&\gamma_1=(f_{1\downarrow}+f_{1\downarrow}^{\dagger})e^{-\int_{0}^{r}\Delta( r^{\prime})/v_2dr^{\prime}}.
\end{aligned}
\end{equation}
 The chiral symmetry $\mathcal{S}$ acts on  $\gamma_1$ as $\mathcal{S}\gamma_1\mathcal{S}^{-1}=\gamma_1$.

In the basis of $\{f_{2}(\bm q),f_{2}^{\dagger}(-\bm q),f_{3}(\bm q),f_{3}^{\dagger}(-\bm q)\}$, the BdG Hamiltonian of describing the Dirac cones centered at $\bar{M}_2$ and $\bar{M}_3$ can be written as 
\begin{widetext}
\begin{equation}
\begin{aligned}
H_{\text{BdG}}=v_2(-1/2q_x\eta_0\tau_zs_y+\sqrt{3}/2q_y\eta_z\tau_zs_y+\sqrt{3}/2q_x\eta_z\tau_0s_x+1/2q_y\eta_0\tau_0s_x)+\Delta_0\tanh (r/\xi)(\frac{x}{r}\eta_0\tau_ys_y-\frac{y}{r}\eta_0\tau_xs_y).
\end{aligned}
\end{equation}    
\end{widetext}
Here $H_{\text{BdG}}$ respects the magnetic mirror symmetry $\mathcal{M_T}=-\eta_xK$, which leads to the chiral symmetry $\mathcal{S}=-\eta_x\tau_x$.
$H_{\text{BdG}}$ is block-diagonal in the $\eta$ space and can be written as

\begin{widetext}
\begin{equation}
\begin{aligned}
&H_{\text{BdG}}=H_2\oplus H_3,\\
&H_2=\left(\begin{matrix}
0 & (\sqrt{3}/2+i/2)v_2q_{-} & 0 & -\Delta_0(\bm r)e^{-i\theta}\\
(\sqrt{3}/2-i/2)v_2q_{+} & 0 & \Delta_0(\bm r)e^{-i\theta} & 0\\
0 & \Delta_0(\bm r)e^{i\theta} & 0 & (\sqrt{3}/2-i/2)v_2q_{+}\\
-\Delta_0(\bm r)e^{i\theta} & 0 & (\sqrt{3}/2+i/2)v_2q_{-} & 0
\end{matrix}\right),\\
&H_3=\left(\begin{matrix}
0 & (-\sqrt{3}/2+i/2)v_2q_{-} & 0 & -\Delta_0(\bm r)e^{-i\theta}\\
(-\sqrt{3}/2-i/2)v_2q_{+} & 0 & \Delta_0(\bm r)e^{-i\theta} & 0\\
0 & \Delta_0(\bm r)e^{i\theta} & 0 & (-\sqrt{3}/2-i/2)v_2q_{+}\\
-\Delta_0(\bm r)e^{i\theta} & 0 & (-\sqrt{3}/2+i/2)v_2q_{-} & 0
\end{matrix}\right),
\end{aligned}
\end{equation}    
\end{widetext}
where $q_{-}-=e^{-i\theta}(q_r-iq_{\theta})$ and $q_{+}=e^{i\theta}(q_r+iq_{\theta})$ with $q_r=-i\partial_r$ and $q_{\theta}=-i/r\partial_{\theta}$. $H_2$ and $H_3$, respectively, host the MZM
\begin{equation}
\begin{aligned}
&\gamma_2=(e^{-i\pi/3}f_{2\downarrow}+e^{i\pi/3}f_{2\downarrow}^{\dagger})e^{-\int_{0}^{r}\Delta( r^{\prime})/vdr^{\prime}},\\
&\gamma_3=(e^{-i2\pi/3}f_{3\downarrow}+e^{i2\pi/3}f_{3\downarrow}^{\dagger})e^{-\int_{0}^{r}\Delta( r^{\prime})/vdr^{\prime}}.
\end{aligned}
\end{equation}
It can be checked that $\mathcal{S}_1\gamma_{2,3}\mathcal{S}_1^{-1}=\gamma_{3,2}$. The three-fold rotational symmetry $\mathcal{C}_3$ is defined as 
\begin{equation}
\begin{aligned}
&\mathcal{C}_3f_1\mathcal{C}_3^{-1}=e^{-i\pi/3}f_2,\\
&\mathcal{C}_3f_2\mathcal{C}_3^{-1}=e^{-i\pi/3}f_3,\\
&\mathcal{C}_3f_3\mathcal{C}_3^{-1}=-e^{-i\pi/3}f_1,\\
\end{aligned}
\end{equation}
where $\mathcal{C}_3^3=1$. Thus, we have 
\begin{equation}
\begin{aligned}
\mathcal{C}_3\gamma_{1,2,3}\mathcal{C}_3^{-1}=\gamma_{2,3,1}.
\end{aligned}
\end{equation}
Thus, in the subspace expanded by $\{\gamma_0,\gamma_1,\gamma_2,\gamma_3\}$, the chiral symmetry $\mathcal{S}$ and three-fold rotational symmetry $\mathcal{C}_3$ are, respectively, represented by
\begin{equation}
\begin{aligned}
\mathcal{S}=\left(\begin{matrix}
1 & 0 & 0 & 0\\
0 & 1 & 0 & 0\\
0 & 0 & 0 & 1\\
0& 0 & 1 & 0
\end{matrix}\right),
\mathcal{C}_3=\left(\begin{matrix}
1 & 0 & 0 & 0\\
0 & 0 & 1 & 0\\
0 & 0 & 0 & 1\\
0& 1 & 0 & 0
\end{matrix}\right).
\end{aligned}
\end{equation}

\bibliography{reference}

\end{document}